\documentclass[aps,floats]{revtex4}
\usepackage{amsmath,amssymb}
\usepackage{graphicx,epsfig}

\begin{document}
\bibliographystyle {plain}

\def\oppropto{\mathop{\propto}} 
\def\opsimeq{\mathop{\simeq}}
\def\opoverderline{\mathop{\overline}}
\def\operarrow{\mathop{\longrightarrow}}
\def\opsim{\mathop{\sim}}

\def\fig#1#2{\includegraphics[height=#1]{#2}}
\def\figx#1#2{\includegraphics[width=#1]{#2}}


\title{ Random field Ising model :  \\
statistical properties of low-energy excitations and of equilibrium avalanches } 


\author{ C\'ecile Monthus and Thomas Garel }
 \affiliation{Institut de Physique Th\'{e}orique, CNRS and CEA Saclay
91191 Gif-sur-Yvette cedex, France}

\begin{abstract}
With respect to usual thermal ferromagnetic transitions, the zero-temperature finite-disorder critical point of the Random-field Ising model (RFIM) has the peculiarity to involve some 'droplet' exponent $\theta$ that enters the generalized hyperscaling relation $2-\alpha= \nu (d-\theta)$. In the present paper, to better understand the meaning of this droplet exponent $\theta$ beyond its role in the thermodynamics, we discuss the statistics of low-energy excitations generated by an imposed single spin-flip with respect to the ground state, as well as the statistics of equilibrium avalanches i.e. the magnetization jumps that occur in the sequence of ground-states as a function of the external magnetic field. The droplet scaling theory predicts that the distribution $dl/l^{1+\theta}$ of the linear-size $l$ of low-energy excitations transforms into the distribution $ds/s^{1+{\theta/d_f}}$ for the size $s$ (number of spins) of excitations of fractal dimension $d_f$ ($s \sim l^{d_f}$). In the non-mean-field region $d<d_c$, droplets are compact $d_f=d$, whereas in the mean-field region $d>d_c$, droplets have a fractal dimension $d_f=2 \theta$ leading to the well-known mean-field result $ds/s^{3/2}$. Zero-field equilibrium avalanches are expected to display the same distribution $ds/s^{1+{\theta/d_f}}$. We also discuss the statistics of equilibrium avalanches integrated over the external field and finite-size behaviors. These expectations are checked numerically for the Dyson hierarchical version of the RFIM, where the droplet exponent $\theta(\sigma)$ can be varied as a function of the effective long-range interaction $J(r) \sim 1/r^{d+\sigma}$ in $d=1$.

\end{abstract}

\maketitle

\section{ Introduction}

Since its introduction by Imry and Ma \cite{imry-ma},
the random field Ising model where Ising spins $S_i = \pm 1$ interact ferromagnetically ($J_{i,j}>0$)
and experience random fields $h_i$ of strength $W$
\begin{eqnarray}
E =  - \sum_{<i,j>} J_{i,j} S_i S_j - \sum_i h_i S_i
\label{defrfim}
\end{eqnarray}
has remained one of the most studied disordered model 
over the years (see for instance the review \cite{nattermann} and references therein).
In addition to the usual critical exponents
$(\alpha,\beta,\gamma,\nu)$ which govern respectively the singularities of the free-energy
$f_{sing} \sim \vert t \vert^{2-\alpha} $,  magnetization $m \sim \vert t \vert^{\beta} $, 
 susceptibility $\gamma \sim \vert t \vert^{\gamma} $, and correlation length 
$\xi \sim \vert t \vert^{-\nu} $ as in thermal ferromagnetic phase transitions, 
the zero-temperature phase transition that occurs at some finite disorder strength $W_c$ in the
Random Field Ising Model (RFIM) has the peculiarity to involve a new exponent called $\theta>0$,
which enters the generalized  hyperscaling relation (see for instance the recent discussion \cite{binder}
and references therein)
\begin{eqnarray}
2-\alpha= \nu (d-\theta)
\label{violationhyperscaling}
\end{eqnarray}
This specificity of the RFIM can be understood as follows.
For usual thermal transitions, the free-energy singularity $F(\xi^d)$
associated to a correlated volume $\xi^d$ is of order unity
\begin{eqnarray}
F(\xi^d) \sim T
\label{freecorr}
\end{eqnarray}
This corresponds to the density singularity
\begin{eqnarray}
f_{sing} \sim \frac{T}{\xi^d} \sim \vert T_c-T \vert^{d \nu} \sim \vert T_c-T \vert^{2-\alpha}
\label{hyperscaling}
\end{eqnarray}
and leads to the usual hyperscaling relation $2-\alpha= d \nu $.
For the RFIM however, 
the singular ground state energy $E(\xi^d)$ of a correlated volume $\xi^d$
is not of order unity as in Eq. \ref{freecorr}, but instead grows as some power of the size $\xi$ 
\begin{eqnarray}
E_{sing}(\xi^d) \sim \xi^{\theta}
\label{xitheta}
\end{eqnarray}
so that the singular energy density is governed instead by
\begin{eqnarray}
e_{sing} \simeq   \frac{\xi^{\theta}}{\xi^d} \simeq
\vert W-W_c \vert^{ \nu (d - \theta) } \sim \vert W-W_c \vert^{2-\alpha}
\label{esingeffxi}
\end{eqnarray}
leading to Eq. \ref{violationhyperscaling}.
As emphasized by Bray and Moore \cite{bray-moore}, the presence of $\theta$ is directly related to the 
zero-temperature nature of the fixed point : for a thermal fixed point occurring at a finite $T_c$,
the fixed point corresponds to a fixed ratio $J_L/T_c$ so that the renormalized coupling $J_L$ is invariant 
under a change of scale $L$; for the RFIM however, the fixed point correspond to a fixed ratio $J_L/W_L$
where $W_L$ represents the renormalized disorder strength. 
So even if this ratio is fixed, both the renormalized coupling $J_L$
and the disorder strength $W_L$ can actually both grow as $L^{\theta}$ with the scale $L$.
This explains how $\theta$ enters the singular part of the energy in Eq. \ref{xitheta}.
Another insight into the physical meaning of $\theta$ comes from the following
scaling picture :  in a correlated volume of size $\xi^d$, the random fields correspond to an averaged
external field of size
\begin{eqnarray}
h_{eff}(\xi^d) \simeq \frac{1}{\xi^d} \sum_{i=1}^{\xi^d} h_i \simeq \xi^{-d/2}
\label{heffxi}
\end{eqnarray}
which is expected to produce a magnetization of order
\begin{eqnarray}
m_{eff}(\xi^d) \simeq \chi h_{eff}(\xi^d) \simeq \chi \xi^{-d/2}
\label{meffxi}
\end{eqnarray}
and to lead to the following singularity in the energy density
\begin{eqnarray}
e_{sing} \simeq m_{eff}(\xi^d) h_{eff}(\xi^d) \simeq \chi \xi^{-d}
\simeq \vert W-W_c \vert^{-\gamma + d \nu}
\label{fsingeff}
\end{eqnarray}
If this analysis is correct, the exponent $\theta$ of Eq. \ref{esingeffxi}
should be directly related to the exponents $\gamma$ and $\nu$
\begin{eqnarray}
\theta = \frac{\gamma}{\nu}  
\label{thetagamma}
\end{eqnarray}
Since this analysis neglects possible correlations, 
Eq. \ref{thetagamma} is usually not considered as true,
but it can be converted into the following bound \cite{sch-sof,nattermann}
\begin{eqnarray}
\theta \geq \frac{\gamma}{\nu}  
\label{thetagammaineq}
\end{eqnarray}
Taking into account the identity $2-\alpha=2 \beta+\gamma$ and the generalized hyperscaling
relation of Eq. \ref{violationhyperscaling}, the inequality \ref{thetagamma} can be rewritten as
\begin{eqnarray}
\theta \geq \frac{d}{2} -  \frac{\beta}{\nu}
\label{thetabetanu}
\end{eqnarray}
Recent works based on nonperturbative functional renormalization \cite{tissier} have concluded
that the equality (\ref{thetagamma}) does not hold in general. However numerically,
it seems difficult to obtain a clear evidence,
since the inequalities (\ref{thetagammaineq}) or (\ref{thetabetanu})
are actually satisfied as egalities within numerical errors
 for the short-range model in various dimensions :
 in dimension $d=3$, the value of $\frac{\beta}{\nu}$
turns out to be extremely small of the order $ \frac{\beta}{\nu}\simeq 0.012$ \cite{middleton3D},
and the droplet exponent $\theta \simeq 1.49$ \cite{middleton3D} is not really optimized
with respect to the non-optimized value $d/2=3/2$. In dimension $d=4$, 
one has a 'reasonable' finite value $ \frac{\beta}{\nu}\simeq 0.19$
\cite{middleton4D} and the droplet exponent $\theta \simeq 1.82$ \cite{middleton4D} remains close
to  $d/2-\frac{\beta}{\nu}$.

Besides this thermodynamic analysis, one expects that the exponent $\theta$ has also the meaning
of a 'droplet exponent' as a consequence of Eq. \ref{xitheta}.
In particular, it should govern the statistics of critical droplets,
defined as low-energy excitations,
as well as the statistics of 'equilibrium avalanches', i.e. the avalanches between ground states as
a function of the external field $H$.
The aim of the present paper is to study in details these statistical properties of droplets
and avalanches. The various predictions are checked via numerics for
  the one-dimensional Dyson hierarchical version,
where large systems can be studied, and where the value of the exponent $\theta$
can be varied as a function of
the exponent $\sigma$ of the long-range ferromagnetic interactions.
Previous studies on these questions for the short-range model in dimension $d=3$
can be found in \cite{alava,hartmann} for the statistics of low-energy excitations,
and in \cite{frontera,liu-dahmen} for the statistics of equilibrium avalanches.

The paper is organized as follows.
After a reminder on the RFIM in the presence of long-range interactions in section \ref{sec-reminder},
we describe the thermodynamical properties of the Dyson hierarchical model in section \ref{sec_dyson}.
In section \ref{sec_droplet}, we discuss the statistics of critical droplets, i.e. of low-energy
excitations generated by an imposed single spin-flip with respect to the true ground-state.
Section \ref{sec_ava} is then devoted to the statistics of equilibrium avalanches, i.e. 
the magnetization jumps that occur in the ground state as the external field is varied.
In section \ref{sec_fsmf}, we discuss the finite-size properties in the mean-field region
$d>d_c$ where usual finite-size scaling is known to break down.
Our conclusions are summarized in section \ref{sec_conclusion}.
In Appendix \ref{app_recursion}, we describe how the sequence of ground states as a function of the external field can be constructed via an exact recursion for the Dyson hierarchical RFIM model.

\section{ Reminder on the RFIM 
with long-range interactions }

\label{sec-reminder}

For statistical-physics models in finite dimension $d$, 
 a phase transition usually exists only above some lower critical dimension $d>d_l$.
 Thermodynamical critical exponents depend upon $d$ in a region $d_l < d < d_c$ 
below the upper critical dimension $d_c$,
and then take their mean-field values for $d>d_c$.
For the short-range RFIM, one expects for instance $d_l=2$ and $d_c=6$.
Besides this short-range case, it is often convenient, both theoretically and numerically,
to consider the case where the ferromagnetic coupling $J_{i,j}$ in Eq. \ref{defrfim}
is long-range and decays only as a power-law in the distance $r=\vert i - j \vert$
\begin{eqnarray}
J(r) \sim \frac{1}{r^{d+\sigma}}
\label{jrdsigma}
\end{eqnarray}
where $\sigma>0$ to have an extensive energy.
Renormalization Group
 analysis of this long-range random-field model can be found in \cite{grinstein,brayLR,weir}.

\subsection{ Imry-Ma argument for the lower critical dimension $d_l$  }

Let us consider the stability of the ferromagnetic ground state where all spins are $(+)$,
with respect to small random fields.
If we flip a domain $v \sim l^d$ spins,
the ferromagnetic cost scales as the double integral of Eq. \ref{jrdsigma}
\begin{eqnarray}
E^{DW}(l)  \propto l^{d-\sigma}
\label{edwcost}
\end{eqnarray}
whereas the random fields may correspond to an energy gain of order
\begin{eqnarray}
E^{RF}(l) =   \sum_{i=1}^{l^d}  h_i \propto W l^{\frac{d}{2}}
\label{edwgain}
\end{eqnarray}
This argument yields that the lower critical dimension is
\begin{eqnarray}
d_l=2 \sigma
\label{dlsigma}
\end{eqnarray}
Indeed for $d>d_l$,  $E^{DW}(l)> E^{RF}(l)$ for sufficiently large $l$,
so the ferromagnetic ground state is stable with respect to
small random fields, and one needs a finite critical disorder $W_c>0$ to destroy it.
On the contrary for  $d<d_l$, $E^{DW}(l) < E^{RF}(l)$ for sufficiently large $l$,
so the ferromagnetic ground state is unstable with respect to small random fields, i.e. $W_c=0$.
In the short-range case, the domain-wall cost scales as the surface $E^{DW}(l)  \propto l^{d-1}$
yielding the usual value $d_l=2$.

\subsection{ Mean-field region  $d>d_c$   }

\label{sec_mfthermo}

The mean-field exponents for the thermodynamical observables at $H=0$
 in the thermodynamic limit $L \to + \infty$ are given by the usual values
\begin{eqnarray}
e_{sing}(W) \sim \vert W-W_c \vert^{2-\alpha_{MF}} \ \ {\rm with } \ \ \alpha_{MF}=0 \nonumber \\
m(W) \sim \vert W-W_c \vert^{\beta_{MF}} \ \ {\rm with } \ \ \beta_{MF}=\frac{1}{2}  \nonumber \\
\chi(W) \sim \vert W-W_c \vert^{-\gamma_{MF}} \ \ {\rm with } \ \ \gamma_{MF}=1
\label{thermomf}
\end{eqnarray}

The Gaussian nature of the fixed point
yields that  the {\it connected } correlation (averaged over disorder)
\begin{eqnarray}
C_{connected}(r) \equiv \overline{ <S_0 S_r >- <S_0>< S_r >} 
\label{correconnected}
\end{eqnarray}
reads for $\sigma \leq 2$
\begin{eqnarray}
C_{connected}(r) \sim  \int d^dq e^{i q r} {\hat C}_{connected}(q)
\ \ {\rm with } \ \  \  {\hat C}_{connected}(q) \sim \frac{T}{\vert W-W_c \vert + \vert q \vert^{\sigma} }
\label{correconnectedgaussian}
\end{eqnarray}
where the term $\vert q \vert^{\sigma}$ represents the leading low-$q$ singularity of
the Fourier transform of the coupling $J(r) \sim 1/r^{d+\sigma}$
(the usual short-range case corresponds the usual term $q^2$  recovered for $\sigma \geq 2$).
The term $\vert W-W_c \vert$ with power unity comes from the compatibility with 
the value $\gamma_{MF}=1$ for the susceptibility since
\begin{eqnarray}
\chi(W) && = \int d^dr C_{connected}(r) \sim  \int d^dr   \int d^dq e^{i q r} {\hat C}_{connected}(q) \nonumber \\ 
&& = {\hat C}_{connected}(q=0) \sim  \vert W-W_c \vert^{- 1 }
\label{ximf}
\end{eqnarray}

The large-$r$ behavior of the connected correlation of Eq. \ref{correconnectedgaussian}
is thus the following power-law at criticality
\begin{eqnarray}
C_{connected}^{W=W_c}(r) \sim  \int d^dq e^{i q r} \frac{T}{ q^{\sigma} }
 \propto r^{\sigma-d} = \frac{1}{r^{d-2+\eta_{MF}}}  \ \ {\rm with } \ \ \eta_{MF} = 2-\sigma
\label{correconnectedgaussiancriti}
\end{eqnarray}
which leads to the following finite-size divergence of the susceptibility at $W_c$
\begin{eqnarray}
\chi_{L}(W_c) = \int^L d^dr C_{connected}^{W=W_c}(r)  = \int^L d^dr   r^{\sigma-d}  \sim L^{\sigma}  
\label{ximffinitesize}
\end{eqnarray}
Off criticality, the exponential decay of the connected correlation defines the correlation
length $\xi$
\begin{eqnarray}
C_{connected}(r) \sim  \int dq e^{i q r} \frac{T}{\vert W-W_c \vert + q^{\sigma} }
 \propto e^{- \frac{r}{\xi}} \ \ {\rm with } \ \ \xi \sim \vert W-W_c \vert^{-\nu_{MF}} \ \ 
{\rm and } \ \ \nu_{MF}=\frac{1}{\sigma}
\label{numf}
\end{eqnarray}

The upper critical dimension $d_c$ is the dimension $d$ where the generalized hyperscaling relation of Eq. 
\ref{violationhyperscaling} is satisfied {\it with the mean-field exponents } 
\begin{eqnarray}
2-\alpha_{MF}= \nu_{MF} (d_c-\theta_{MF})
\label{violationhyperscalingdc}
\end{eqnarray}
where the mean-field exponent $\theta_{MF}$ is given by Eq. \ref{thetagamma}
\begin{eqnarray}
\theta_{MF}= \frac{\gamma_{MF}}{\nu_{MF}} = \frac{1}{\nu_{MF}}
\label{thetamf}
\end{eqnarray}
leading to
\begin{eqnarray}
d_c= \frac{3}{ \nu_{MF} }
\label{dcgene}
\end{eqnarray}
For $\sigma \leq 2$ where $\nu_{MF}=1/\sigma $ (Eq. \ref{numf}), this yields
\begin{eqnarray}
d_c(\sigma \leq 2)= 3 \sigma
\label{dcsigma}
\end{eqnarray}
whereas the usual short-range value $d_c=6$ is recovered for $\sigma \geq 2$ where $\nu_{MF}=1/2$
and $\theta_{MF}=2 $.

\section{ Dyson hierarchical model with random fields }

\label{sec_dyson}

To better understand the notion of phase transition in statistical physics,
Dyson \cite{dyson} has introduced long ago a hierarchical ferromagnetic spin model,
which can be studied via exact renormalization for probability distributions.
In this approach, the hierarchical couplings are chosen to mimic
  effective long-range power-law couplings in real space,
so that phase transitions are possible already in one dimension.
This type of hierarchical model has thus attracted a great interest 
in statistical physics, both among mathematicians
\cite{bleher,gallavotti,book,jona} and among physicists \cite{baker,mcguire,Kim,Kim77}.
In the field of quenched disordered models, Dyson hierarchical models have been 
introduced for spin systems with random fields \cite{randomfield}
or with random couplings \cite{sgdysonAT,sgdysonHS,sgdysonR},
as well as for Anderson localization \cite{bovier,molchanov,krit,kuttruf,fyodorov,EBetOG,
fyodorovbis,us_dysonloc}. In the following, we consider the Dyson hierarchical random field model
 \cite{randomfield}.

\subsection{ Definition of the model }

The Hamiltonian for $2^N$ spins in an exterior magnetic field $H$ reads
\begin{eqnarray}
{\cal H}_{2^N}(H;\{h_1,...,h_{2^N}\}) && \equiv - \sum_{i=1}^{2^N} (H+h_i) S_i 
 \label{hamilton} \\
  && - J_1 \left[ (S_1+S_2)^2 +(S_3+S_4)^2 + (S_5+S_6)^2+(S_7+S_8)^2 + ... \right]
\nonumber \\
&& - J_2 \left[ (S_1+S_2+S_3+S_4)^2 + (S_5+S_6+S_7+S_8)^2 + ...
\right] \nonumber \\
&& - J_3 \left[ (S_1+S_2+S_3+S_4+S_5+S_6+S_7+S_8)^2 + ...
\right] - .... \nonumber \\
&& - J_N \left(S_1+S_2+...+S_{2^N-1}+S_{2^N}\right)^2 \nonumber
\end{eqnarray}
The random fields $h_i$ represent quenched disordered variables drawn with some distribution,
 for instance the box distribution of width $W$
\begin{eqnarray}
P_W^{Box}(h) = \frac{1}{W} \theta \left( - \frac{W}{2} \leq h \leq \frac{W}{2} \right)
\label{box} 
\end{eqnarray}
The ferromagnetic couplings $J_n$ are chosen to decay exponentially with the level
$n$ of the hierarchy
\begin{eqnarray}
J_n =  \left( \frac{1}{2^{1+\sigma}} \right)^{n}
\label{jn} 
\end{eqnarray}
To make the link with the physics of long-range one-dimensional models,
it is convenient to consider that the sites $i$ of the Dyson model
 are displayed on a one-dimensional lattice, with a lattice spacing unity.
Then the site $i=1$ is coupled via the coupling $J_n$ to 
the sites $2^{n-1} < i \leq 2^n$. At the scaling level, the hierarchical model
is thus somewhat equivalent to the
following power-law dependence in the real-space distance $L_n=2^n$
\begin{eqnarray}
J_n =  \left( \frac{1}{L_n} \right)^{1+\sigma}
\label{jnspace} 
\end{eqnarray}
 One thus expects that the Dyson hierarchical model will have
the same essential properties as the long-range model discussed
in the previous section for the case $d=1$.
In particular, the model is well defined with an extensive energy for $\sigma>0$.
The lower critical dimension $d_l=2 \sigma$ (Eq. \ref{dlsigma}) coming from the Imry-Ma argument
and the upper critical dimension $d_c=3 \sigma$ (Eq. \ref{dcsigma}) coming from the analysis
of the Gaussian fixed point are the same : so in $d=1$, the mean-field region 
and the non-mean-field region exist respectively in the following domains of the parameter $\sigma$
\begin{eqnarray}
&& {\rm mean-field \ region } \ \ d=1>d_c \ \ {\rm for } \ \ 0<\sigma < \frac{1}{3} \nonumber \\
&& {\rm non-mean-field \ region } \ \ d_l<d=1<d_c \ \ {\rm for } \ \  \frac{1}{3}<\sigma < \frac{1}{2}
\label{mfnonmf} 
\end{eqnarray}

\subsection{ Value of the droplet exponent  $\theta=\sigma$ in the non-mean-field region }

Whereas for the short-range model, we are not aware of any conjecture concerning 
the value of $\theta$ in the non-mean-field region $d_l=2<d<d_c=6$,
Grinstein \cite{grinstein} has conjectured that in the presence of long-range interactions,
the exponent $\theta$ always takes the following simple value 
\begin{eqnarray}
\theta=\sigma
\label{conjecture}
\end{eqnarray}
which is consistent with various limits (in particular $d \to d_l=2 \sigma$ and $d \to d_c=3 \sigma$)
and various perturbative expansions \cite{grinstein,brayLR,weir}.
This conjecture was then shown to be wrong perturbatively at order $O(\epsilon^2)$ \cite{brayLR}
for the long-ranged case discussed in section \ref{sec-reminder},
but to be true in the Dyson hierarchical version \cite{randomfield}.

Eq \ref{conjecture} can be understood via the following scaling analysis.
Let us introduce the exponent $y$ governing the power-law decay of the magnetization per spin
with the system size $L$ exactly at the critical point $W_c$  
\begin{eqnarray}
m(W_c,L) \oppropto_{L \to +\infty} L^{-y}
\label{mcritiL}
\end{eqnarray}
The effective magnetic field per spin resulting from the random fields scales as
\begin{eqnarray}
h^{eff}_L = \frac{1}{L^d} \sum_{i=1}^{L^d}  h_i \oppropto_{L \to +\infty} L^{- \frac{d}{2}}
\label{heffCLT}
\end{eqnarray}
The characteristic Zeeman energy of the $L^d$ spins of magnetization $M_L=L^d m(W_c,L)$ is thus
\begin{eqnarray}
E_Z(W_c,L) \simeq h^{eff}_L M_L \oppropto_{L \to +\infty} L^{\frac{d}{2}-y}
\label{ezcritiL}
\end{eqnarray}
whereas the ferromagnetic energy associated to the effective coupling at scale $L$ 
(see Eq. \ref {jrdsigma})
\begin{eqnarray}
J^{eff}_L \sim L^{-d-\sigma}
\label{jferroeff}
\end{eqnarray}
behaves as
\begin{eqnarray}
E_{ferro}(W_c,L) \simeq J^{eff}_L M_L^2 \oppropto_{L \to +\infty} L^{d-\sigma-2y}
\label{eferrocritiL}
\end{eqnarray}

At the critical point, the two energies of Eq. \ref{ezcritiL} and Eq \ref{eferrocritiL}
should remain in competition at all scales,
i.e. they should have the same scaling
\begin{eqnarray}
y= \frac{d}{2}-\sigma
\label{ymcritiL}
\end{eqnarray}
Then, the two energies scale as
\begin{eqnarray}
E_Z(W_c,L) \sim E_{ferro}(W_c,L) \sim L^{\theta} \ \ {\rm with} \ \ \theta= \sigma
\label{ethetacritiL}
\end{eqnarray}
So Eq. \ref{conjecture} relies on the fact that the renormalized
ferromagnetic coupling $ J^{eff}_L$ is directly given by the power-law defining the model
(Eq. \ref{jferroeff}) as a  consequence of the exact hierarchical structure.
As a final remark, let us mention that 
in the short-range case, the effective renormalized ferromagnetic coupling $ J^{eff}_L$
cannot be simply estimated, and this is why there is no simple conjecture
for the value of $\theta$.

\subsection{ Finite-size properties of thermodynamical observables }

As a consequence of  Eq. \ref{conjecture} that holds exactly for the Dyson hierarchical model,
 many finite-size behaviors exactly at criticality
can be explicitly computed as a function of $\sigma$, for all $d>d_l$,
i.e. both in the non-mean field region $d<d_c$ and in the mean-field region $d>d_c$
(since in the mean-field region, one has also $\theta_{MF}=\sigma$ (Eqs \ref{numf} and \ref{thetamf} )). 
In particular, 
the singular part of the energy density scales as (Eq. \ref{ethetacritiL})
\begin{eqnarray}
e_{sing}(W_c,L) \sim \frac{L^{\theta}}{L^d} = L^{\sigma-d}
\label{esingcritiL}
\end{eqnarray}
the divergence of the susceptibility scales as  (Eqs \ref{mcritiL} \ref{heffCLT} and \ref{ymcritiL})
\begin{eqnarray}
\chi(W_c,L) \sim \frac{m(W_c,L)}{h^{eff}_L} \sim L^{d/2-y} = L^{\sigma}
\label{chicritiL}
\end{eqnarray}
The exponent of the two-point correlation may also be obtained as
\begin{eqnarray}
\overline{S_i S_{i+r} } \sim m^2(W_c,r) \sim \frac{1}{ r^{2 y} } = 
\frac{1}{r^{d-2+{\tilde \eta}}} 
 \ \ {\rm with } \ \ 
{\tilde \eta}  = 2-2 \sigma 
\label{correcritiL}
\end{eqnarray}
At low temperature $T$, the {\it connected } correlation involves another exponent
\begin{eqnarray}
\overline{ <S_0 S_r >- <S_0>< S_r >} \simeq (T) r^{- (d-2+\eta)} 
\label{correconnectedcritiL}
\end{eqnarray}
As a consequence of the scaling of Eq. \ref{ethetacritiL}, only a rare fraction
$T/r^{\theta}$ can contribute to the connected correlation function,
 so that one expects the following shift 
\begin{eqnarray}
 \eta=\tilde{\eta} + \theta =  2- \sigma 
\label{etaetabar}
\end{eqnarray}

\subsection{ Numerical results on the magnetization }

As explained in Appendix \ref{app_recursion}, the hierarchical structure of the Dyson model
allows to write exact recursions to compute the ground states that occur as a function of the
external field for each disordered sample. 
We have studied the following sizes $2^6 \leq L \leq 2^{21}$
 with a corresponding statistics of $4.10^7  > n_s(L) \geq 45.10^3 $ independent samples.

\subsubsection{ Location of the critical point }

\begin{figure}[htbp]
 \includegraphics[height=6cm]{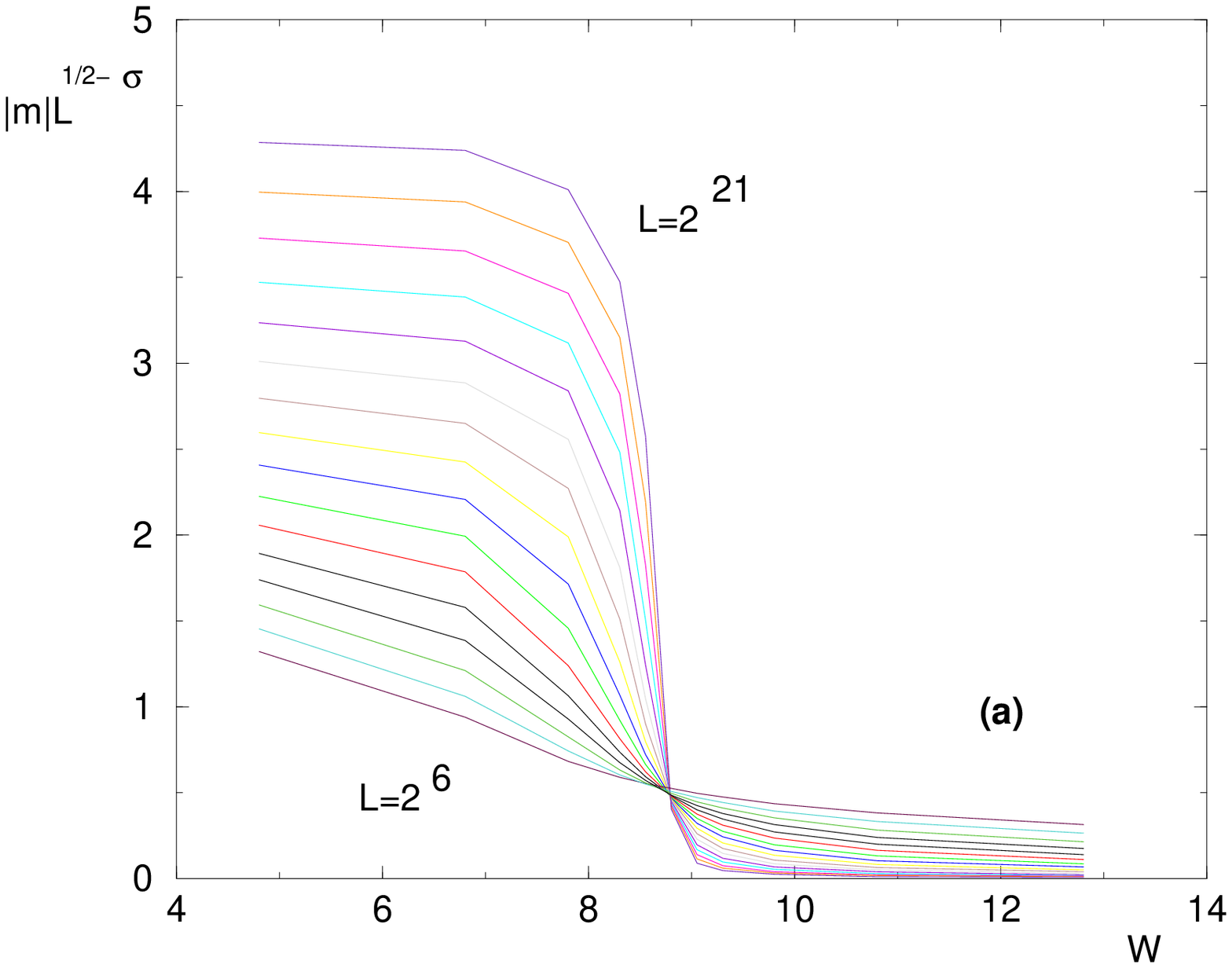} 
\hspace{2cm}
 \includegraphics[height=6cm]{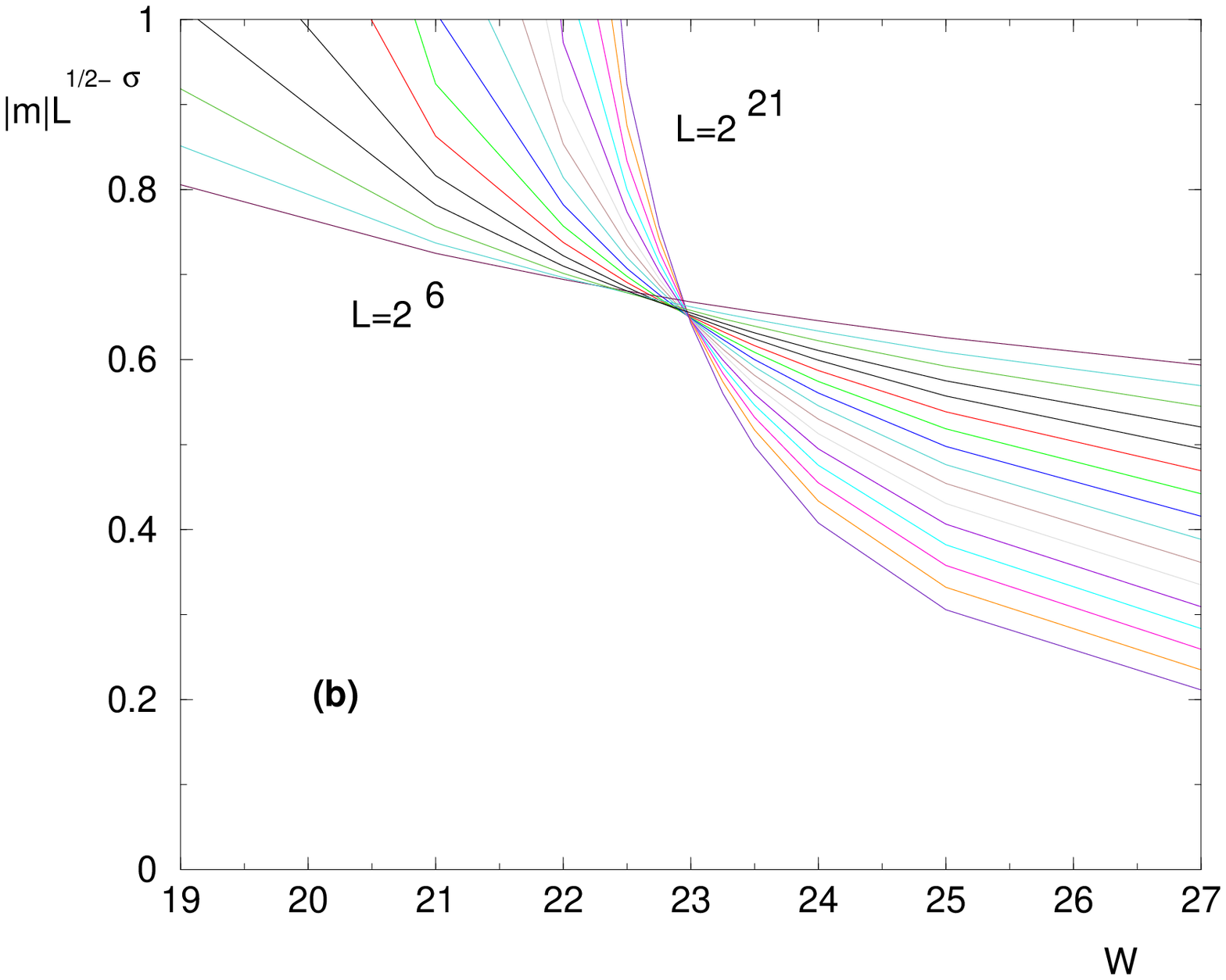}
\vspace{1cm}
\caption{ Location of the critical point $W_c$ via the crossing of the rescaled
magnetization curves $L^{1/2-\sigma} \vert m(W,L) \vert$ as a function of the disorder strength $W$,
for sizes $2^6 \leq L \leq 2^{21}$.
(a) Case $\sigma=0.4$ (Non-mean-field region)  : the curves $L^{0.1} \vert m(W,L) \vert $ cross near $W_c \simeq 8.8$
(b) Case $\sigma=0.2$ (Mean-field region): the curves $L^{0.3} \vert m(W,L)\vert $ cross near $W_c \simeq 23$.  }
\label{figcrossing}
\end{figure}

Exactly at criticality, the magnetization $\vert m(W_c,L) \vert $ is expected to decay as $L^{-y}$
with $y=1/2-\sigma$ (see Eqs \ref{mcritiL} and \ref{ymcritiL}), both in the mean-field region
and in the non-mean-field region.
With our numerical data,
we indeed find that the curves $L^{1/2-\sigma} m(W,L)$ for various sizes $L$
cross more and more sharply as $L$ grows : this crossing allows to locate the critical disorder
strength $W_c$, as shown on Fig \ref{figcrossing} for the two cases $\sigma=0.2$ and $\sigma=0.4$.
We have also data concerning the cases $\sigma=0.1$ and $\sigma=0.3$ (not shown).
All numerical results presented below concern the critical point $W_c$, i.e. with our data
\begin{eqnarray}
W_c(\sigma=0.1) && \simeq  50 \nonumber \\
W_c(\sigma=0.2) &&\simeq  23 \nonumber \\
W_c(\sigma=0.3) && \simeq  13.5 \nonumber \\
W_c(\sigma=0.4) && \simeq  8.8
\label{wcsigma}
\end{eqnarray}

\begin{figure}[htbp]
 \includegraphics[height=6cm]{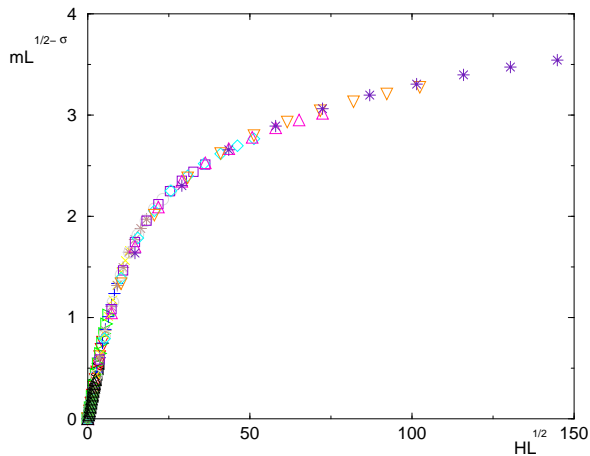}
\vspace{1cm}
\caption{ Finite-size scaling of the magnetization at $W_c$
 as a function of the external magnetic field $H$ :
$L^{1/2-\sigma} m(H,L)$ as a function of the scaling variable $H L^{1/2}$
(see  Eq. \ref{fssminh}) 
 for $\sigma=0.4$ (non-mean-field region).
 }
\label{figfss54sigma=0.4}
\end{figure}

\subsubsection{ 'Non-mean-field' region $\frac{1}{3} < \sigma < \frac{1}{2} $ with usual finite-size scaling  }

In the 'non-mean-field' region, thermodynamical 
observables follow standard finite-size scaling forms in $(W-W_c)L^{1/\nu}$
involving the correlation length exponent $\nu$,
so that the known behaviors in $L$ at criticality for the magnetization, 
susceptibility and singular energy yield 
\begin{eqnarray}
\frac{\beta}{\nu} && =y= \frac{1}{2}-\sigma \nonumber \\
\frac{\gamma}{\nu} &&  =\sigma \nonumber \\
\frac{2-\alpha}{\nu}&& =1-\sigma
\label{ratiosfss}
\end{eqnarray}
Since the correlation length $\nu$ is not known exactly, the values of $(\alpha,\beta,\gamma)$
are not known either, but the ratios of Eq. \ref{ratiosfss} are exactly known.

Besides these good finite-size properties in $(W-W_c)L^{1/\nu}$ at $H=0$, one also expects good
finite-size properties in the variable $H L^{1/2}$ (see Eq. \ref{heffCLT}) exactly at $W_c$
The following finite-size scaling form
\begin{eqnarray}
\overline{  m(W_c,H \geq 0)  } \propto L^{\sigma-1/2 } \ \
  {\cal M} \left( H L^{\frac{1}{2}} \right)
\label{fssminh}
\end{eqnarray}
is indeed well-satisfied 
 as shown on Fig. \ref{figfss54sigma=0.4}  for the case $\sigma=0.4$.

The finite-size properties in the mean-field region $d>d_c$ 
where usual finite-size scaling is known to break down will be discussed later
in section \ref{sec_fsmf}.

In conclusion of this section, the Dyson hierarchical RFIM is a convenient model 
where large systems can be studied numerically with a good statistics, and where the droplet exponent
$\theta=\sigma$ can be varied by choosing the parameter $\sigma$ of the effective long-range interactions.
Now that we have located the zero-temperature finite-disorder critical point for various $\sigma$,
we may study the statistics of low-energy excitations and of equilibrium avalanches in the 
remaining sections.

\section{ Statistics of low-energy excitations at criticality }

\label{sec_droplet}

In disordered systems, there can be states that have an energy
very close to the ground state energy but which are very different
from the ground state in configuration space.
 In the droplet theory, developed initially for the spin-glass phase \cite{Fis_Hus_SG}
and then for the frozen phase of the directed polymer in a random medium 
\cite{Fis_Hus,Hwa_Fis,us_dropletDP},
the low-temperature physics is described
in terms of rare regions with nearly degenerate excitations
which appear with a probability that decays with a power-law of their size.
In these models of spin-glasses or directed polymers, the droplet exponent $\theta$
is a property of the low-temperature disorder-dominated phase $T<T_c$.
In the RFIM however, the originality is that the droplet exponent $\theta$
is a property of the zero-temperature finite-disorder critical point.
We have already recalled in the introduction how this exponent $\theta$ enters in the critical
properties of thermodynamical observables, in particular in the generalized hyperscaling relation
of Eq. \ref{violationhyperscaling}. In the present section, we discuss how this droplet exponent $\theta$
governs the power-law distribution of low-energy excitations at $W_c$.

\subsection{ Low-energy excitations generated by an imposed spin-flip }

To generate low-energy excitations above the ground state in disordered systems,
various procedures have been followed in the literature
(see for instance the various methods concerning spin-glasses 
\cite{kawashima,lamarcq,berthier,picco,h_moore}).
In the following, since we wish to generate an elementary local excitation,
we have chosen the 'single spin flip method', already used in Ref. \cite{hartmann} 
for the short-range 3D RFIM. The idea is the following : in each disordered sample,
we first compute the true ground-state. Then we impose the flip of a given spin $S_{i_0}$ with respect
to its orientation in the true ground state, and we compute the new modified ground state when this
constraint is taken into account. We measure the number $s$ of spins that are different
in the two ground states. This excitation has a finite-energy by construction : 
if only $S_{i_0}$ flips, the cost is simply $\Delta E= 2 \vert h_{i_0}^{loc} \vert$
 in terms of the local field $ h_{i_0}^{loc}$ ; if the systems chooses to flip $s$ spins, it is because
the energy cost is lower.

At criticality, the probability distribution $D^{W_c}_L(s) $
of the number $s$ of spins of this low-energy excitation 
is expected to decay as a power-law in the thermodynamic limit $L \to +\infty$
\begin{eqnarray}
D^{W_c}_L(s) \opsimeq_{L \to +\infty}  \frac{1}{s^{\tau_D}}
\label{dropletpower}
\end{eqnarray}
where $\tau_D$ is defined by this equation.
For finite $L$, a finite cut-off 
\begin{eqnarray}
s_*(L) \oppropto_{L \to +\infty}  L^{\rho}
\label{cutoffrhodroplet}
\end{eqnarray}
 is expected to govern the far-exponential decay
\begin{eqnarray}
D^{W_c}_L(s)  \opsimeq_{s \to +\infty} 
e^{- \frac{s }{s_*(L)} }\
\label{fssdroplet}
\end{eqnarray}
Let us first discuss the relation between the exponent  $\tau_D$ of Eq. \ref{dropletpower}
and the droplet exponent $\theta$.

\subsection{ Relation with the statistics of the linear size $l$ of droplets }

From the definition of the droplet exponent $\theta$,
 a droplet of linear size $l$ has an energy cost of order
\begin{eqnarray}
E^{droplet}(l) \propto  l^{\theta} u 
\label{edroplet}
\end{eqnarray}
where $u$ is a positive random variable of order $O(1)$ distributed with some law $p(u)$.
The random variable $u$ is expected to have a zero weight $p(u=0)>0$ at the origin.
The probability to have a droplet of linear size $l$ and of finite energy $E^{droplet}(l)< E_0$
then scales as $Prob(u< \frac{E_0}{l^{\theta}}) \simeq p(u=0) \frac{E_0}{l^{\theta}}$.
Taking into account the logarithmic measure $dl/l$ in the size $l$ to insure independent
droplets, one obtains the following distribution for the linear size $l$
\begin{eqnarray}
dl P^{droplet}(l) \simeq \frac{dl }{l^{1+\theta} } E_0 p(u=0) 
\label{pdroplet}
\end{eqnarray}
as in other models like spin-glasses \cite{Fis_Hus_SG}
or directed polymers \cite{Fis_Hus,Hwa_Fis,us_dropletDP}.

To obtain the statistics in the size $s$ (number of spins) of droplets,
one needs to know the fractal dimension $d_f$ of droplets
\begin{eqnarray}
s \sim l^{d_f}
\label{defdfdroplet}
\end{eqnarray}
The probability distribution in $l$ of Eq. \ref{pdroplet} transforms into the following
power-law distribution of Eq. \ref{dropletpower} with
\begin{eqnarray}
\tau_D=1+ \frac{  \theta}{d_f}  
\label{taudropletthetadf}
\end{eqnarray}

In the context of spin-glasses, a similar relation between the droplet exponent $\theta$
and the avalanche exponent $\tau$ (which coincides 
with the exponent $\tau_D$ of low-energy excitations of Eq. \ref{taudropletthetadf}
as discussed in the next section) 
has been discussed in Ref. \cite{MFavaeq}, where it is assumed that 
the density of droplets is $(dl/l^{1+\theta}) \times l^{d-d_f}$ which is different from Eq.
\ref{pdroplet} whenever the droplets are non-compact $d_f \ne d$. 
We believe that Eq. \ref{pdroplet} is correct, and is consistent with our numerical results
where we generate droplets by flipping {\it an arbitrary spin }, both in the case of
compact or fractal droplets, as we now discuss.

\subsection{ Statistics of the size $s$ of low-energy excitations in the non-mean-field region $d<d_c$ }

\begin{figure}[htbp]
 \includegraphics[height=6cm]{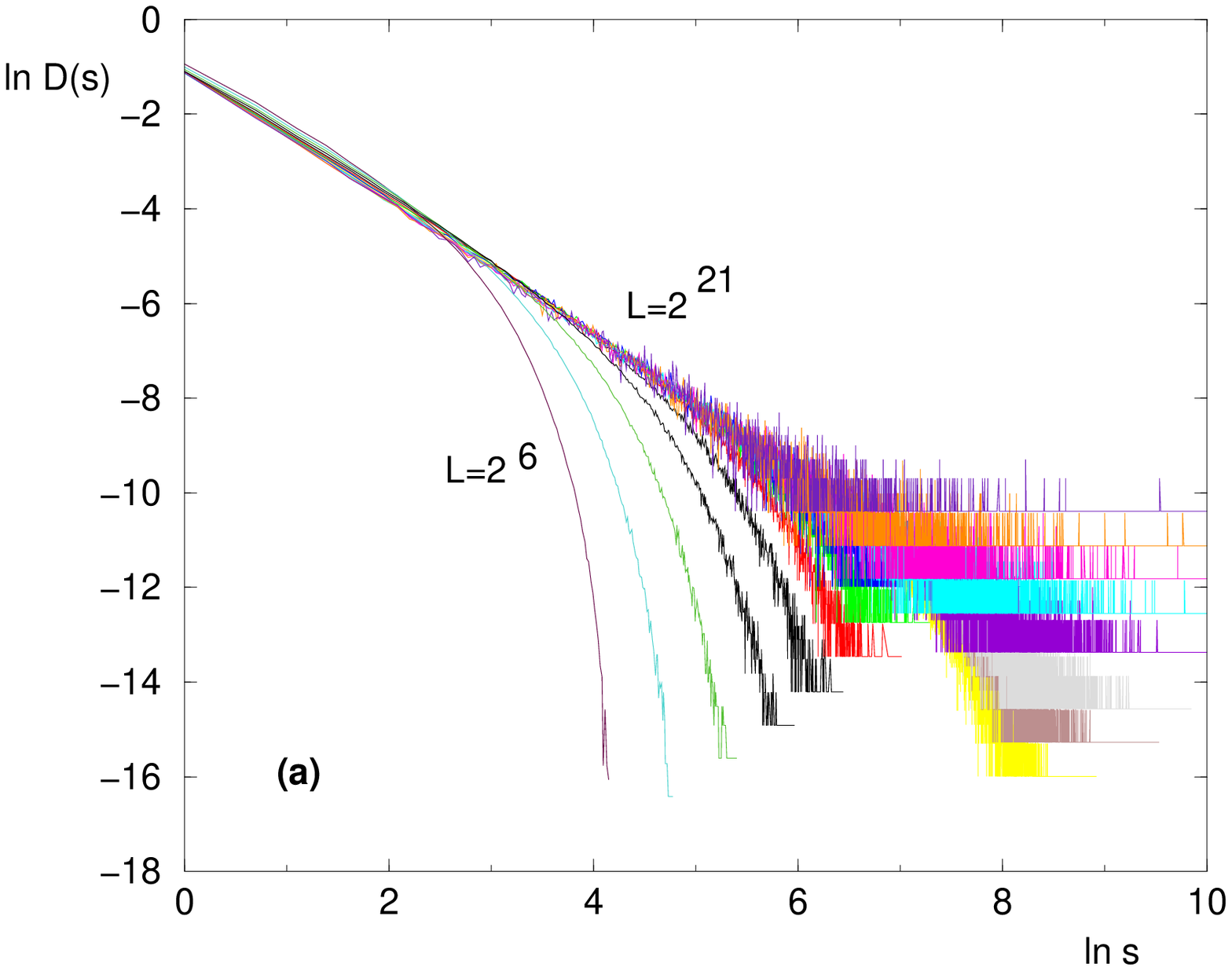}
\hspace{2cm}
 \includegraphics[height=6cm]{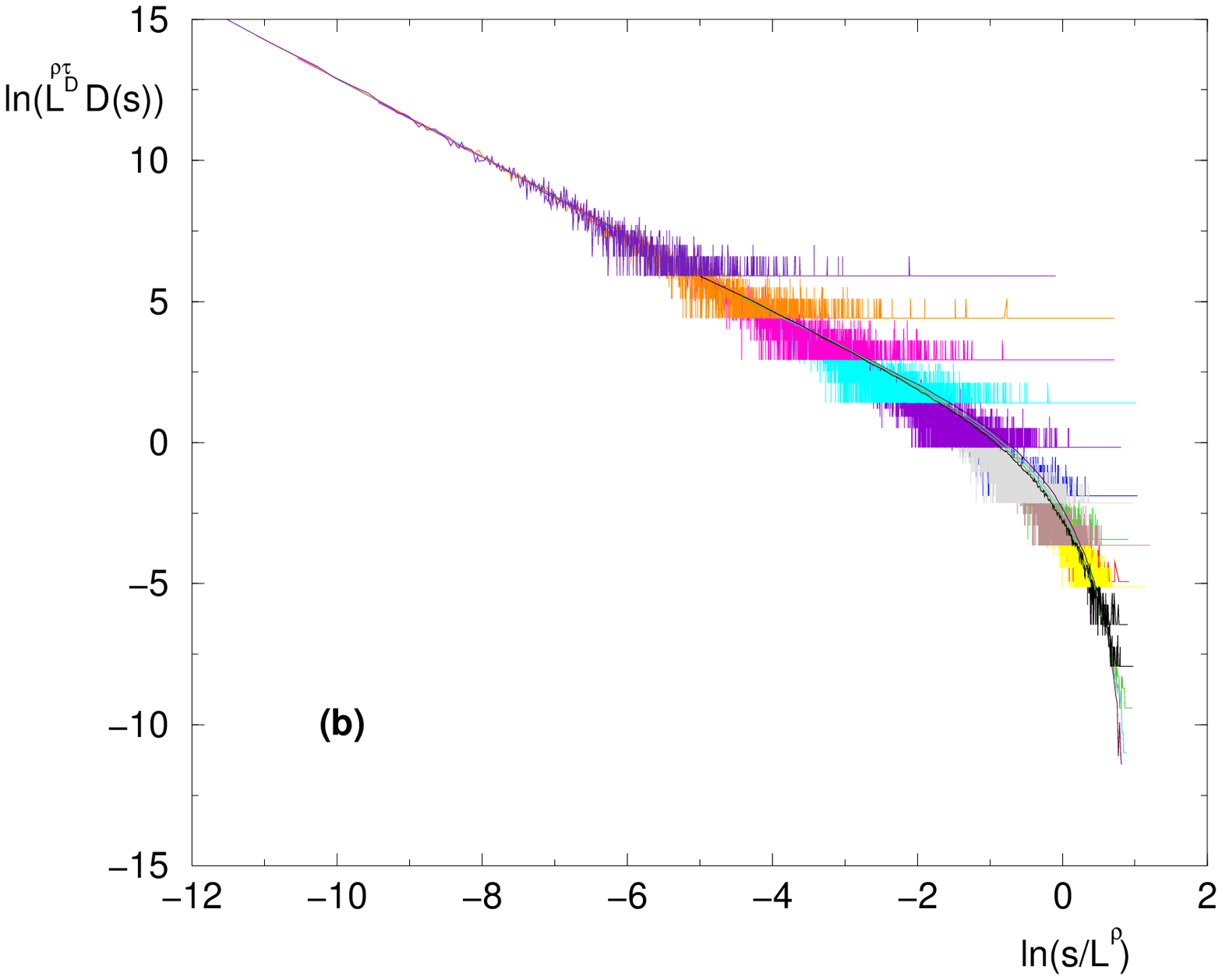}
\vspace{1cm}
\caption{ Distribution $D^{W_c}_L(s)$ of the size $s$ (number of spins)
of critical droplet for $\sigma=0.4$ (non-mean-field region) :
(a) $\ln D^{W_c}_L(s)$ as a function of $\ln s$ to measure the droplet exponent $\tau_D$ in Eq. \ref{dropletpower} :  the slope corresponds to the exponent $\tau_D =1+\theta= 1.4$ 
(b) Finite-size scaling 
$\ln \left( L^{\rho \tau_D} D^{W_c}_L(s) \right)$ as a function of $\ln \left(\frac{s}{L^{\rho}} \right)$
with $\rho=2 \sigma=0.8$. }
\label{figdropletnonMF}
\end{figure}

In the non-mean-field region, one expects that droplets are compact,
i.e. the fractal dimension of Eq. \ref{defdfdroplet} is simply 
\begin{eqnarray}
d_f=d
\label{dfdropletcompact}
\end{eqnarray}
so that Eq. \ref{taudropletthetadf} reads
\begin{eqnarray}
\tau_D=1+  \frac{\theta}{d}   \ \ {\rm for \ \ } d \leq d_c 
\label{taudropletthetaMF}
\end{eqnarray}
By considering avalanches in the next section \ref{sec_ava} (see Eq. \ref{rhofss}),
we expect that the cut-off exponent $\rho$ is directly related to the droplet exponent $\theta$
\begin{eqnarray}
\rho = 2 \theta
\label{rhothetanonMF}
\end{eqnarray}

Our numerical data for the Dyson hierarchical Dyson 
model of parameter $\sigma=0.4$ are in agreement with these prediction.
On Fig. \ref{figdropletnonMF} (a), we show a log-log plot of the droplet distribution $D^{W_c}_L(s)$
for various $L$ and we measure the slope $\tau_D(\sigma=0.4) =1+\theta=1+\sigma= 1.4 $.
On Fig.\ref{figdropletnonMF} (b), we show that a satisfactory data collapse can be obtained in terms of the rescaled variable $ \frac{s}{L^{\rho}}$ with $\rho(\sigma=0.4) = 2\theta=2 \sigma= 0.8$.

\subsection{ Statistics of the size $s$ of low-energy excitations in the mean-field region $d>d_c$ }

\label{sec_mfdroplet}

\begin{figure}[htbp]
 \includegraphics[height=6cm]{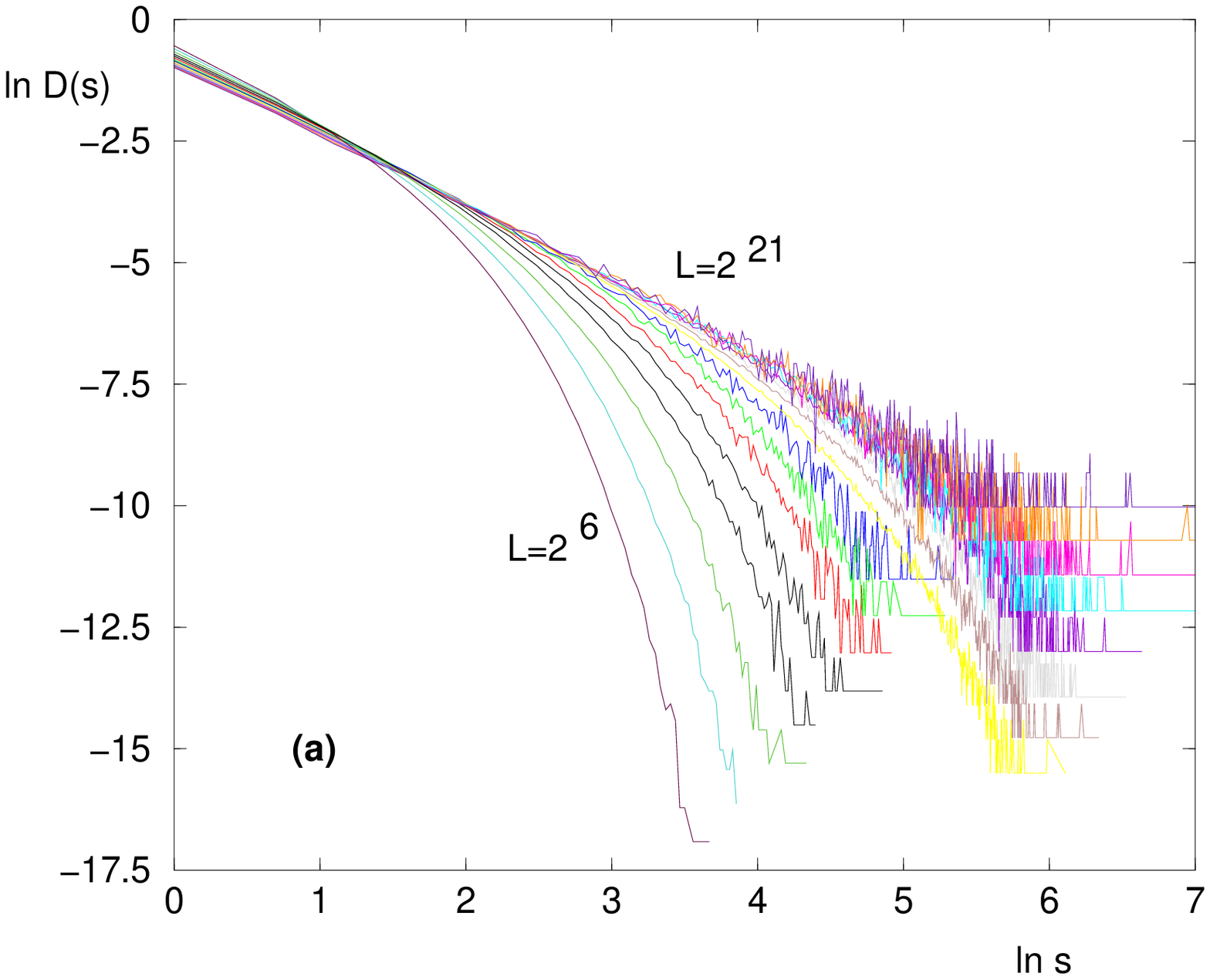}
\hspace{2cm}
 \includegraphics[height=6cm]{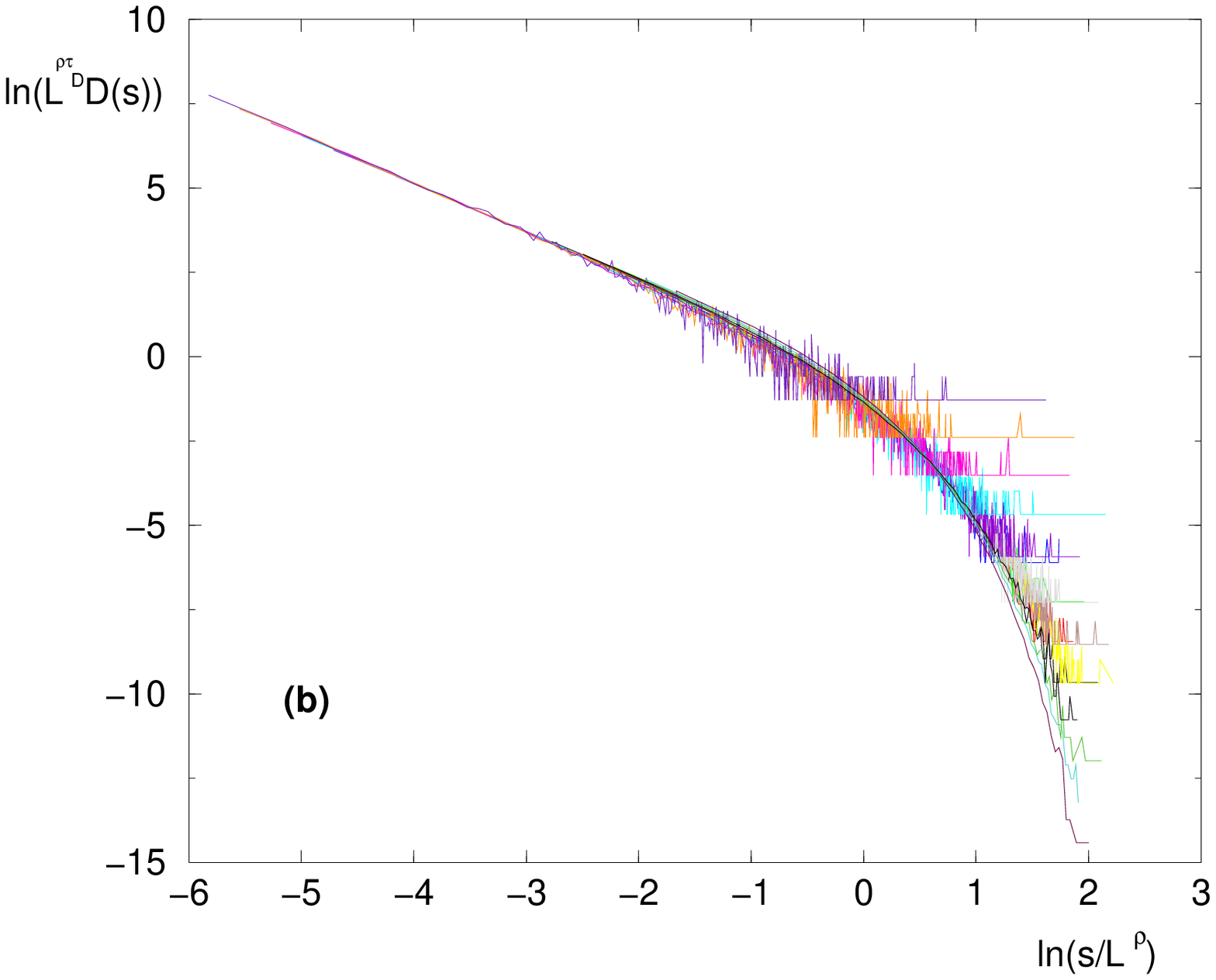}
\vspace{1cm}
\caption{
 Distribution $D^{W_c}_L(s)$ of the size $s$ (number of spins)
of critical droplet for $\sigma=0.2$ (mean-field region) :
(a) $\ln D^{W_c}_L(s)$ as a function of $\ln s$ to measure the droplet exponent $\tau_D$ in Eq. \ref{dropletpower} :  the slope corresponds to the mean-field exponent $\tau_D =\tau_D^{MF}=3/2$ 
(b) Finite-size scaling 
$\ln \left( L^{\rho \tau_D^{MF}} D^{W_c}_L(s) \right)$ as a function of $\ln \left( \frac{s}{L^{\rho}} \right)$ 
with $\rho =2 \sigma=0.4 $. }
\label{figdropletsigma=0.2}
\end{figure}

In the mean-field region $d>d_c$, one expects that 'loops' are not important, 
so that the probability distribution $D(s)$ to return $s$ spins
to obtain the new ground state when one imposes the flip of one given spin $S_{i_0}$
should take the same form as the probability that a spin belongs
to a cluster of size $s$ for the percolation problem of the Bethe Lattice
that has no loops \cite{book_perco}
\begin{eqnarray}
D_{MF}(s) = \frac{1}{s^{\tau_{MF}}} e^{- s (W-W_c)^2 } \ \ {\rm with } \ \ \tau_{MF}=\frac{3}{2}
\label{dropletMF}
\end{eqnarray}
This means that the density $n(s)$ of clusters of size $s$ per spin reads 
\begin{eqnarray}
n(s) = \frac{D_{MF}(s)}{s} = \frac{1}{s^{\tau_{MF}+1}} e^{- s (W-W_c)^2 } 
\label{ns}
\end{eqnarray}
Let us now adapt the Coniglio scaling analysis \cite{coniglio} concerning the percolation 
transition for $d>d_c$ to our present case.
In a volume $l^d$, the number $N_{l^d} $ of clusters has
 the following singularity in terms of the cut-off $s_{max} \sim  (W-W_c)^{-2}$ of Eq. \ref{ns}
\begin{eqnarray}
N_{l^d}  = l^d \sum_{s} n(s) =l^d [ s_{max} ]^{- \tau_{MF}}
 = l^d (W-W_c)^{ 2 \tau_{MF}} 
\label{nl}
\end{eqnarray}
In particular, on a correlated volume $\xi^d = \vert W-W_c \vert^{- \nu_{MF}}$, the number of clusters 
has for singularity
\begin{eqnarray}
N_{\xi^d}  = \xi^d (W-W_c)^{ 2 \tau_{MF}} = \xi^{d-d_c} 
\ \ {with} \ \ d_c= \frac{2 \tau_{MF}}{\nu_{MF}}= \frac{3}{\nu_{MF} } 
\label{nxi}
\end{eqnarray}
As stressed by Coniglio \cite{coniglio}, the meaning of the upper critical dimension $d_c$ 
is that it separates :

(i) the mean-field region $d>d_c$ where the number $N_{\xi^d}$ of clusters 
diverges with $\xi$ as $\xi^{d-d_c} $

(ii)  the non-mean-field region $d<d_c$ where a correlated volume contains a single relevant cluster.

In our present long-range case where $\nu_{MF}=1/\sigma$ (Eq. \ref{numf}) we recover $d_c=3 \sigma$
(Eq \ref{dcsigma}),
whereas the short-range case where $\nu_{MF}=1/2$ corresponds to $d_c=6$.

In the mean-field region, the fact that a correlated volume contains a large number of clusters 
$  N_{\xi^d} \gg 1$ is possible  because each cluster has a fractal dimension $d_f$ given by
\begin{eqnarray}
 s_{max} \sim (W-W_c)^{-2} \sim \xi^{d_f} 
\ \ {\rm with } \ \ d_f=\frac{2 }{\nu_{MF}} = 2 \theta_{MF}
\label{mfdf}
\end{eqnarray}
in agreement with the relation of Eq. \ref{taudropletthetadf}
\begin{eqnarray}
\tau_{MF} =1+ \frac{  \theta_{MF}}{d_f}  = \frac{3}{2}
\label{taudropletthetadfMF}
\end{eqnarray}
In our present long-range case where $\nu_{MF}=1/\sigma$ (Eq. \ref{numf}), the fractal dimension
of droplets is thus
\begin{eqnarray}
 d_f=\frac{2 }{\nu_{MF}}=2 \sigma
\label{df2sigma}
\end{eqnarray}
whereas in the short-range case where $\nu_{MF}=1/2$, the fractal dimension is $d_f=4$.

In this mean-field region where droplets are non-compact, the only constraint
on the maximal linear size of droplets is the system-size $L$ itself : $l_{max} \sim L$,
so we may expect that the maximal size in $s$ of avalanches in a finite system is
\begin{eqnarray}
s_*(L) \sim L^{d_f} = L^{2 \sigma}
\label{srhodfMF}
\end{eqnarray}
Our conclusion is that in the mean-field region, the exponent $\rho$ of the finite-size cut-off
of Eq. \ref{cutoffrhodroplet} should be
\begin{eqnarray}
\rho = d_f = 2 \sigma
\label{rhodfMF}
\end{eqnarray}

Our data for $\sigma=0.1$, $\sigma=0.2$ and $\sigma=0.3$ are indeed compatible with the values
$\tau_{MF}=3/2$ and $\rho=2 \sigma$, as shown on Fig. \ref{figdropletsigma=0.2} for $\sigma=0.2$.

\section{ Statistics of equilibrium avalanches }

\label{sec_ava}

Power-law avalanches occur in many domains of physics. 
In the field of disordered systems, {\it non-equilibrium avalanches} have been much studied,
in particular in the context of driven elastic manifolds in random media 
(see the review \cite{reviewDSF} and references therein, as well as the more recent works
\cite{drivenparticle,depinningav}) and in the Random Field Ising model
(see  \cite{dahmen_sethna,perkovic,review_sethna} and references therein).
To better understand the properties of these non-equilibrium avalanches,
it seems useful to make the comparison with the {\it equilibrium avalanches }
that occur in the same systems. The statistics of equilibrium avalanches has been studied
 for elastic manifolds in random media \cite{staticavainterfaces}, for spin-glasses \cite{MFavaeq}
and for the RFIM \cite{frontera,liu-dahmen}, where some universality has been found
between equilibrium and non-equilibrium avalanches \cite{liu-dahmen}.
In this section, we discuss the statistics of equilibrium avalanches in the RFIM.

\subsection{ Observables concerning equilibrium avalanches }

Let us  introduce the average number of avalanches of size $s$ that occur at $H$ in a system of size $L$
\begin{eqnarray}
N_L (s,H) = \overline{ \sum_i \delta(H-h_{flip}(i)) \ \ \delta(s- \frac{m_{i+1}-m_i}{2}) }
\label{numberaval}
\end{eqnarray}
where the $h_{flip}(i)$ and the $m_i$ are the fields and the magnetization 
occurring in the sequence of ground-states of a sample as a function of the external field
(see more details in Appendix \ref{app_recursion}).
The total number of avalanches for $-\infty < H < +\infty$ is then
\begin{eqnarray}
N_L^{tot}  \equiv \int_{-\infty}^{+\infty} dH \int_0^{+\infty} ds N_L (s,H)
\label{numberavaltot}
\end{eqnarray}
Since during the history, each spin of the $L^d$ spins
has to flip exactly once, one has the exact sum rule
\begin{eqnarray}
 \int_{-\infty}^{+\infty} dH \int_0^{+\infty} ds s N_L (s,H) = L^d
\label{sumrule}
\end{eqnarray}
In terms of these avalanches, the difference between the magnetization at $H$ and at $H=0$
can be written as
\begin{eqnarray}
M(H)-M(0) = 2  \int_{0}^{H} dh \int_0^{+\infty} ds s N_L (s,h) 
\label{deltaMwithaval}
\end{eqnarray}
so that the susceptibility reads
\begin{eqnarray}
\chi_L(H) = \frac{ dm }{ dH} = \frac{1}{L^d} \frac{dM}{dH}  =  \frac{1}{L^d} \int_0^{+\infty} ds s N_L (s,H) 
\label{chiaval}
\end{eqnarray}

\subsection{ Probability distribution of zero-field avalanches at $W_c$ }

\begin{figure}[htbp]
 \includegraphics[height=6cm]{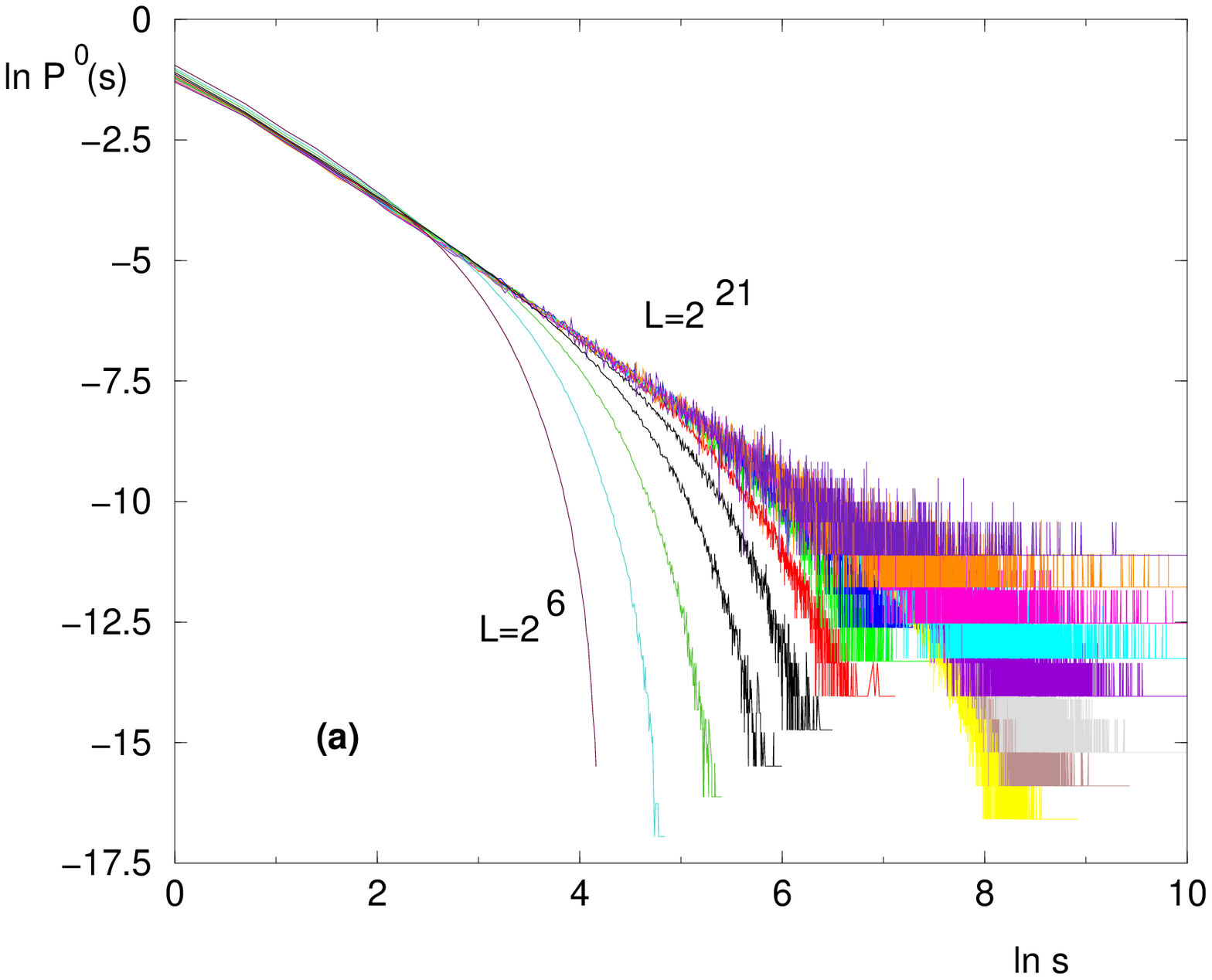}
\hspace{2cm}
 \includegraphics[height=6cm]{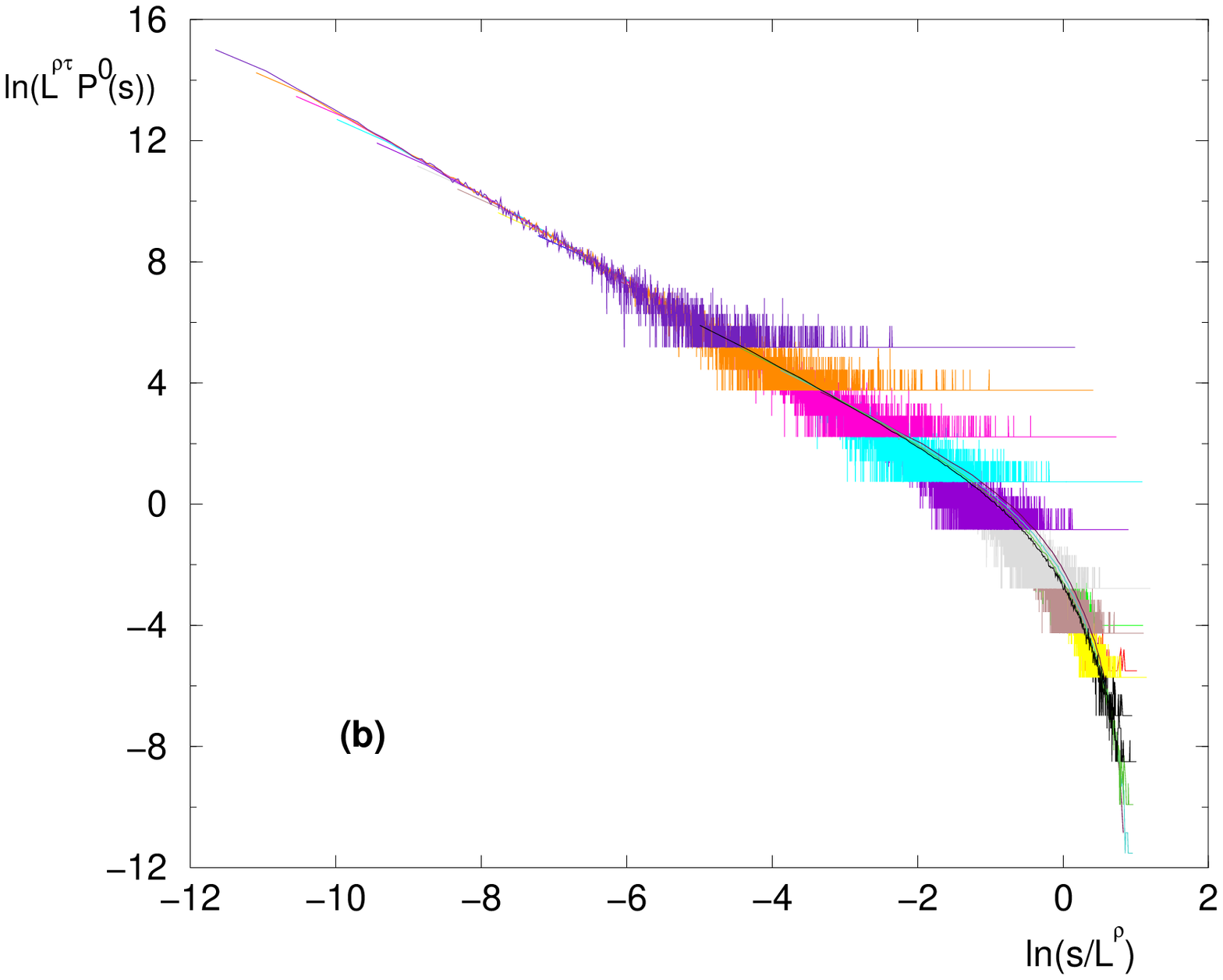}
\vspace{1cm}
\caption{ Probability distribution $P_L^{(H=0)}(s)$
of the size $s$ of avalanches at $W_c$ and $H=0$ in the non-mean-field region for $\sigma=0.4$
and the sizes ($2^6 \leq L \leq 2^{21}$) : 
(a)  $\ln P_L^{(H=0)}(s)$ as a function of $\ln s$  :  the slope corresponds to  $\tau =1+\sigma= 1.4$
(b)  Finite-size scaling analysis :
$\ln \left( L^{\rho \tau} P^{(H=0)}_L(s) \right)$ as a function of $\ln \left( \frac{s}{L^{\rho}} \right) $  
  with $\tau = 1.4$ and $\rho =2 \sigma= 0.8 $.  }
\label{figPzerosigma=0.4}
\end{figure}

\begin{figure}[htbp]
 \includegraphics[height=6cm]{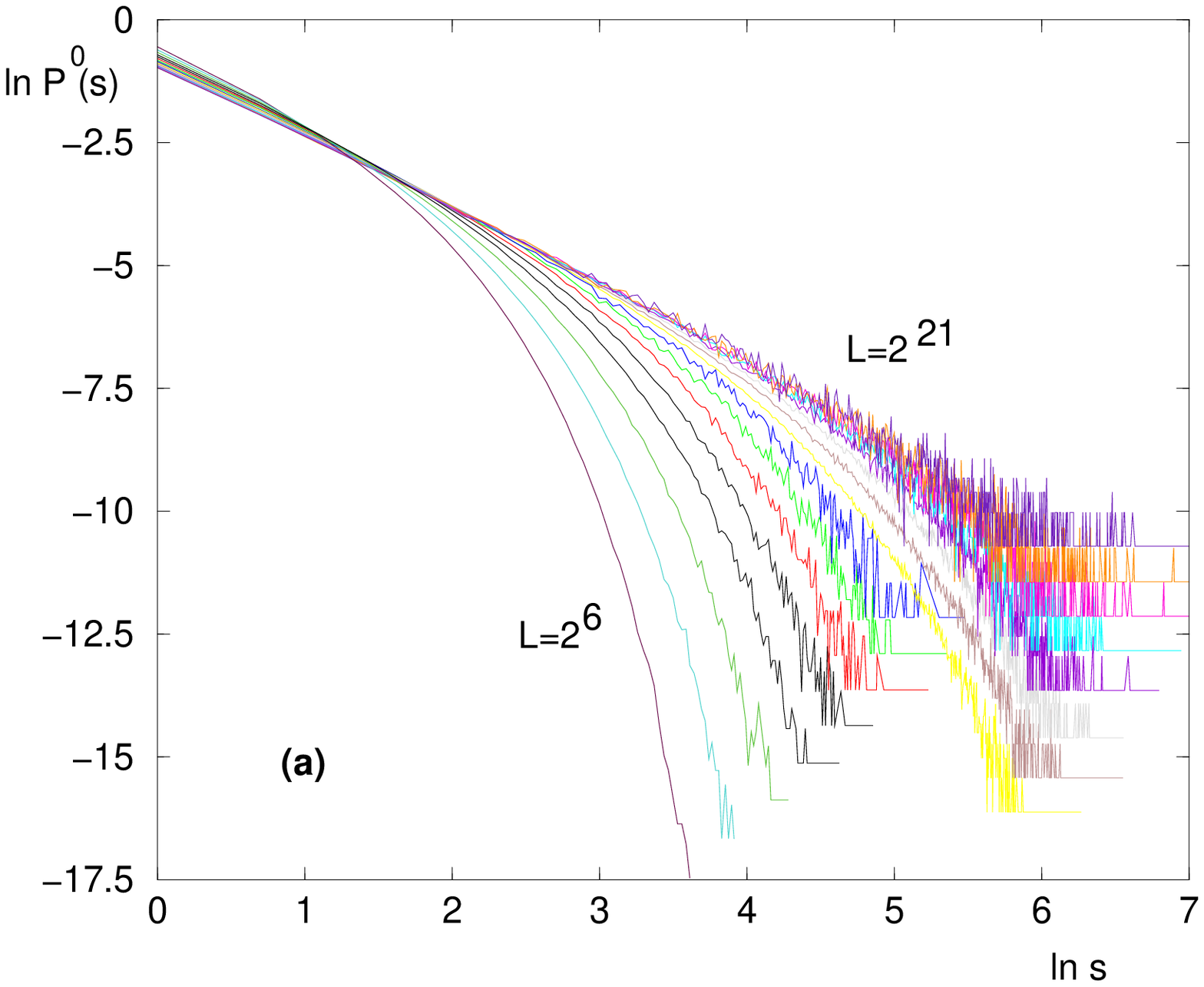}
\hspace{2cm}
 \includegraphics[height=6cm]{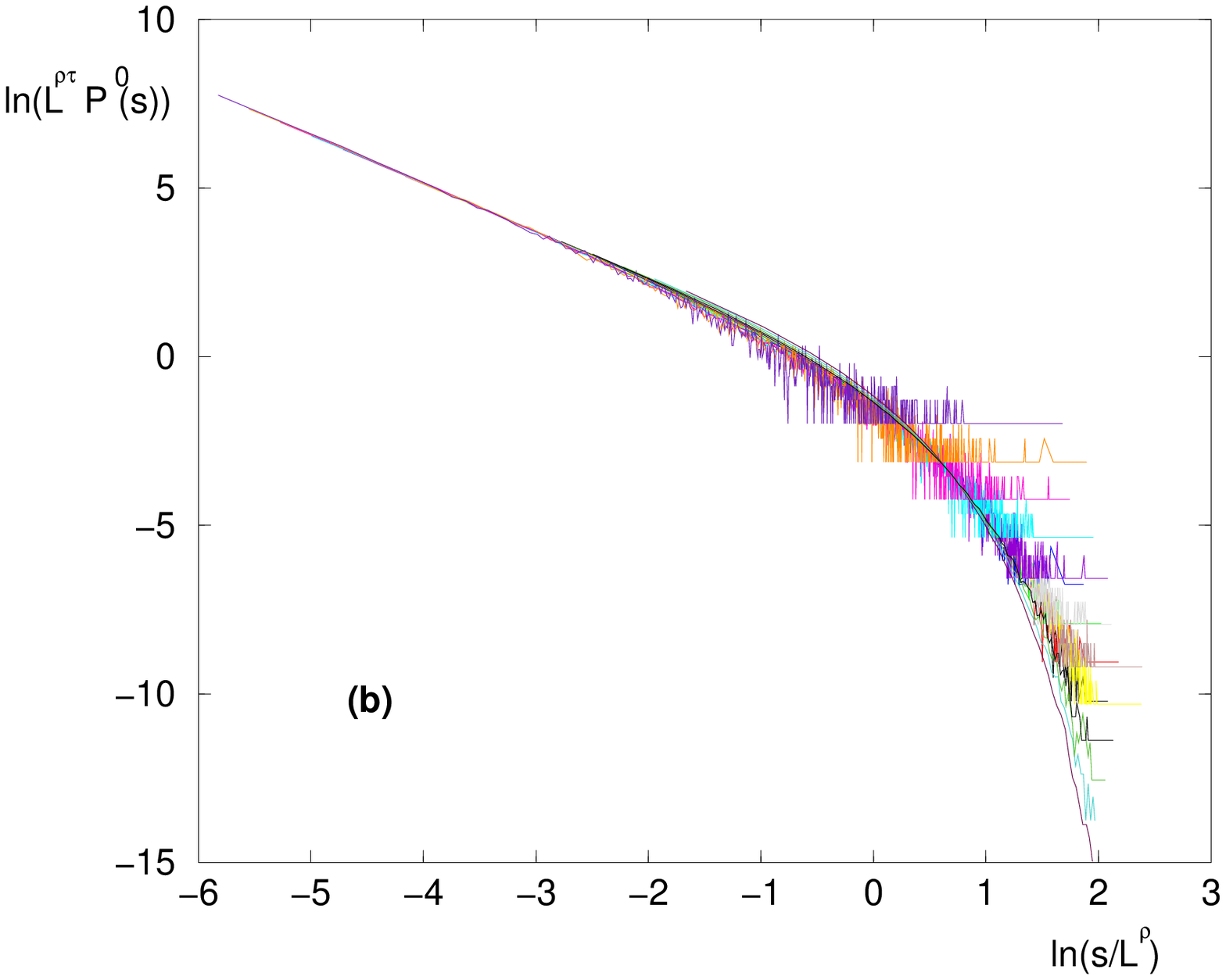}
\vspace{1cm}
\caption{ 
 Probability distribution $P_L^{(H=0)}(s)$
of the size $s$ of avalanches at $W_c$ and $H=0$ in the mean-field region for $\sigma=0.2$
and the sizes ($2^6 \leq L \leq 2^{21}$) : 
(a)  $\ln P_L^{(H=0)}(s)$ as a function of $\ln s$  :
the slope corresponds to the exponent $\tau =\tau_{MF}= \frac{3}{2} $
(b) 
 Finite-size scaling of the probability distribution $\ln P^{(H=0)}_L(s)$ of avalanches :
$\ln \left( L^{\rho \tau} P^{(H=0)}_L(s) \right)$ as a function of $\ln \left( \frac{s}{L^{\rho}} \right)$ 
 with $\tau = \frac{3}{2}  $ and $\rho =2 \sigma= 0.4 $ .
  }
\label{figPzerosigma=0.2}
\end{figure}

To measure the probability distribution of the size $s$ (number of spins) of avalanches exactly at $H=0$
(see the definition of Eq. \ref{numberaval})
\begin{eqnarray}
P_L^{(H=0)}(s) \equiv \frac{  N_L (s,H=0) }{ \int_0^{+\infty} ds N_L (s,H=0) }
\label{Pzero}
\end{eqnarray}
we have considered, in each disordered sample, the avalanche occuring at the smallest $\vert h_{flip}(i)
\vert$.
\begin{eqnarray}
h_{flip}^{min} = {\rm min} \vert h_{flip}(i) \vert
\label{hflipmin}
\end{eqnarray}

The probability distribution $R(h_{loc})$ 
of the local fields $h_{loc}(i) \geq 0$ of spins $S_i$ in the ground state 
is expected to have a finite weight $R(h_{loc}=0)>0$ at the origin $h_{loc}=0$.
Then the closest avalanche from $H=0$  occurs at a field of order 
\begin{eqnarray}
h_{flip}^{min} \sim \frac{1}{L^d}
\label{hflipminL}
\end{eqnarray}
(If one draws $L^d$ variables from this distribution, 
the minimal local field will scale as $h_{loc}^{min} \propto 1/L^d$ 
from the estimate
\begin{eqnarray}
\frac{1}{L^d} = \int_0^{h_{loc}^{min}} dh R(h) = R(0) h_{loc}^{min} 
\label{hlocminL}
\end{eqnarray}
We have checked the validity of Eq. \ref{hflipminL} for the Dyson hierarchical RFIM,
both in the non-mean-field and in the mean-field regions.

When the external field $H$ reaches this $h_{loc}^{min} \propto 1/L^d$, at least one spin becomes unstable,
and it induces an avalanche of size $s$ with some probability $P_L(s)$.
From this argument, it seems natural to expect that this $P_L(s)$ exactly
 coincides with the droplet distribution $D_L(s)$
discussed in the previous section \ref{sec_droplet}
 (the only difference is that for the droplet distribution,
we have chosen to force the flipping of an arbitrary spin, whereas here the field forces the flipping
of the spin having the lowest local field)
\begin{eqnarray}
P_L^{(H=0)}(s) = D_L(s) = \frac{1}{s^{\tau}} e^{- \frac{s}{s_*(L)}}  \ \ {\rm with } \ \ \tau=\tau_D \ \  
\label{Pzerodroplet}
\end{eqnarray}
with the same finite-size cut-off $s_*(L) \sim L^{\rho}$.
As explained in section \ref{sec_droplet}, the difference between the mean-field and the
non-mean-field  regions is that
\begin{eqnarray}
\tau \left( d<d_c \right) && 
=\tau_D\left( d<d_c \right) =1+\frac{\theta}{d} \ \    \ \nonumber \\
 \tau \left( d>d_c \right) && =\tau_D^{MF} =\frac{3}{2} 
\label{Pzerodroplettau}
\end{eqnarray}

Our numerical data for the Dyson hierarchical RFIM of parameter $\sigma=0.4$ (non-mean-field region) 
and $\sigma=0.2$ (mean-field region) are in agreement with this picture, as shown 
on Figures \ref{figPzerosigma=0.4} and \ref{figPzerosigma=0.2} respectively.

\subsection{ Integrated probability distribution of avalanches for $-\infty < H < +\infty$ at $W_c$}

\begin{figure}[htbp]
 \includegraphics[height=6cm]{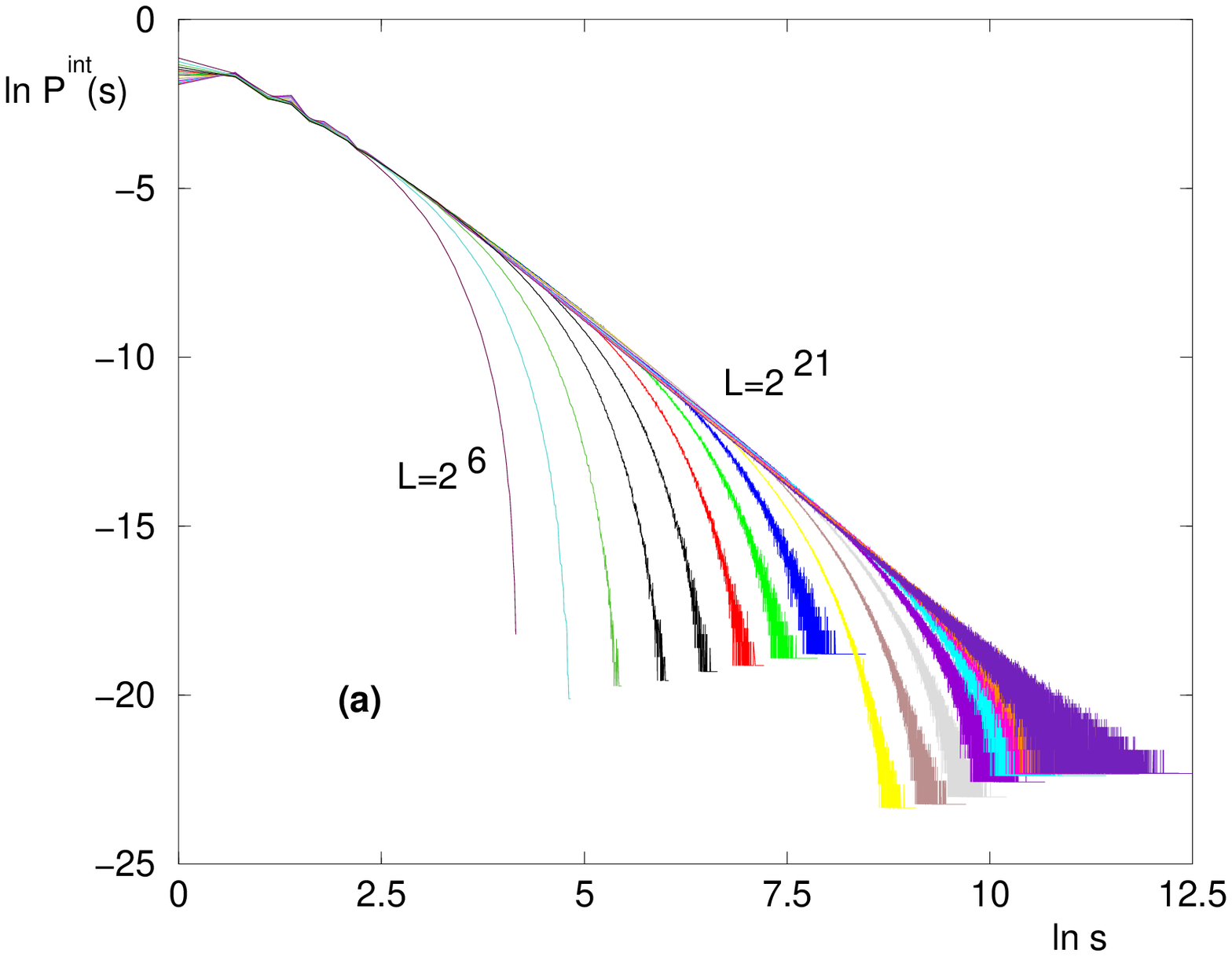}
\hspace{2cm}
 \includegraphics[height=6cm]{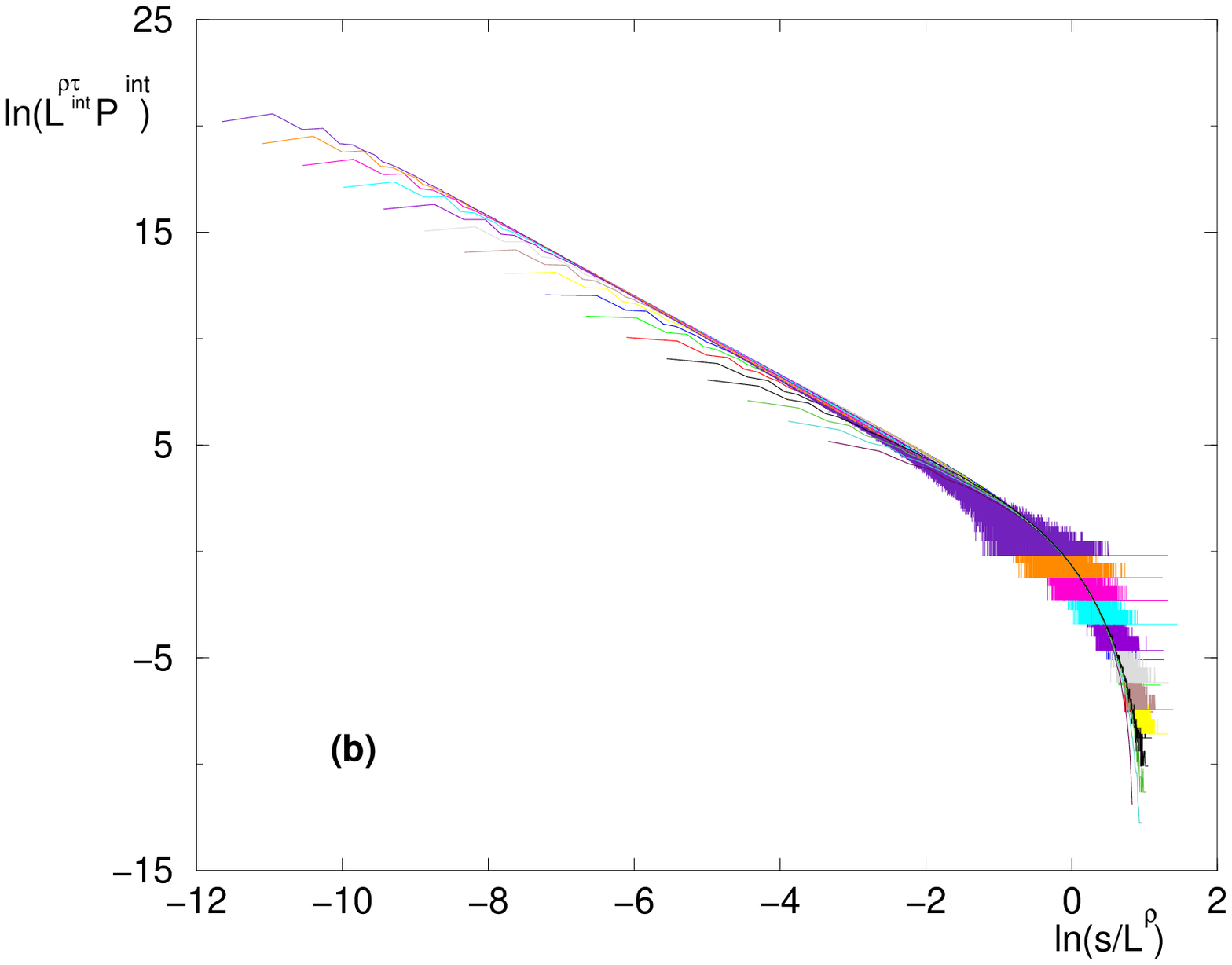}
\vspace{1cm}
\caption{ Integrated distribution $P_L^{int}(s)$
of the size $s$ of avalanches for $-\infty < H < +\infty$
at $W_c$ in the non-mean-field region for $\sigma=0.4$
(sizes $2^6 \leq L \leq 2^{21}$) 
(a)  $\ln P_L^{int}(s)$ as a function of $\ln s$  : 
 the slope corresponds to the exponent $\tau_{int} = \frac{3}{2}+\sigma =  1.9$
(b)  Finite-size scaling analysis :
$\ln \left( L^{\rho \tau_{int}} P^{int}_L(s) \right)$ as a function of $\ln \left( \frac{s}{L^{\rho}} \right)$ 
  with $\tau_{int} = 1.9 $ and $\rho =2 \sigma = 0.8 $.    }
\label{figQintsigma=0.4}
\end{figure}

We have also measured numerically the so-called
 'integrated' distribution of the size $s$ of avalanches
 (where 'integrated' means 'integrated over the external field $-\infty < H < +\infty$'),
\begin{eqnarray}
P_L^{int}(s) \equiv \frac{\int_{-\infty}^{+\infty} dH  N_L (s,H) }{N_L^{tot}}
\label{Qint}
\end{eqnarray}
where the notations $ N_L (s,H) $ and $ N_L^{tot}$ have been introduced in Eqs \ref{numberaval}
and \ref{numberavaltot}.

Lets us now focus on the non-mean-field region $d<d_c$, where 
 one expects that the statistics of avalanches of Eq. \ref{numberaval}
will follow some finite-size scaling form
\begin{eqnarray}
dH ds N_L (s,H) \simeq  dH L^a \frac{ds}{s^{\tau}} \Phi \left( \frac{s}{L^{\rho}}, H s^{\psi}  \right)
\label{fssnav}
\end{eqnarray}
The exponents $\tau$ and $\rho$ have been already introduced
as the exponents that characterize the statistics of zero-field avalanches of Eq. \ref{Pzero}
\begin{eqnarray}
P_L^{(H=0)}(s) \equiv \frac{  N_L (s,H=0) }{ \int_0^{+\infty} ds N_L (s,H=0) }
= \frac{  N_L (s,H=0) }{ \int_0^{+\infty} ds N_L (s,H=0) } 
= \frac{1}{s^{\tau}} \Phi \left( \frac{s}{L^{\rho}}, 0  \right)
\label{Pzeroscaling}
\end{eqnarray}
The exponent $a$ governs the total number $N_L^{tot} $ of avalanches (Eq. \ref{numberavaltot})
occuring in a sample of size $L$
\begin{eqnarray}
N_L^{tot} = \int dH \int_0^{+\infty} ds N_L (s,H) \simeq   L^a 
\label{nltotla}
\end{eqnarray}
The exponent $\psi$ describes the finite-size scaling in the external field $H$.
In this non-mean-field region where avalanches are compact $s \sim l^d$, we expect that 
the appropriate scaling variable is $H l^{d/2} = H s^{1/2}$
so that $\psi$ takes the simple value
\begin{eqnarray}
\psi=\frac{1}{2} \ \ {\rm for } \ \ d<d_c
\label{psinonMF}
\end{eqnarray}

The integrated distribution of avalanches introduced in Eq. \ref{Qint}
can be now written in terms of the finite-size scaling form of Eq. \ref{fssnav}
\begin{eqnarray}
P_L^{int}(s) = \frac{\int_{-\infty}^{+\infty} dH  N_L (s,H) }{N_L^{tot}}
= \frac{1}{s^{\tau}} \int_{-\infty}^{+\infty} dH  \phi \left( \frac{s}{L^{\rho}}, H s^{\psi}  \right)
= \frac{1}{s^{\tau+\psi}} \int_{-\infty}^{+\infty} dh  \phi \left( \frac{s}{L^{\rho}}, h  \right)
\label{Qintfss}
\end{eqnarray}
In the thermodynamic limit $L \to +\infty$, the integrated distribution is also a power-law
\begin{eqnarray}
P_{L=+\infty}^{int}(s) \sim \frac{1}{s^{\tau_{int}}} 
\label{Qintthermo}
\end{eqnarray}
where the exponent $\tau_{int}$ is shifted from the exponent $\tau=1+\frac{\theta}{d}$
of the zero-field avalanches (Eq. \ref{Pzerodroplet}) by the factor $\psi_{nonMF}=1/2$ (Eq. \ref{psinonMF} )
\begin{eqnarray}
\tau_{int} = \tau + \psi = \frac{3}{2}+\frac{\theta}{d}
\label{relationtauinttau}
\end{eqnarray}
Our numerical data for $\sigma=0.4$ are in agreement with this value $\tau_{int}=\frac{3}{2}+\sigma
=1.9$ as shown on Fig. \ref{figQintsigma=0.4} (a).

Let us now write the consistency equations that fixes the values of $\rho$ and $a$.
Eq. \ref{chiaval} concerning the susceptibility yields
\begin{eqnarray}
\chi_L(H=0) =  \frac{1}{L^d} \int_0^{+\infty} ds \  s N_L (s,H=0) 
=  L^{a-d}  \int_0^{+\infty} ds \  s^{1-\tau}  \Phi \left( \frac{s}{L^{\rho}}, 0 \right)
\sim L^{a-d} \left( L^{\rho} \right)^{2-\tau}
\label{chiavalfssava}
\end{eqnarray}
From the divergence of the susceptibility at criticality $\chi_L \sim L^{\frac{\gamma}{\nu}} = L^{\theta} $
(Eq. \ref{thetagamma}), we obtain the relation
 \begin{eqnarray}
\theta =a-d +\rho (2-\tau)
\label{relationchiaval}
\end{eqnarray}

The other consistency relation comes from the sum rule of Eq. \ref{sumrule} that reads
\begin{eqnarray}
L^d && = \int_{-\infty}^{+\infty} dH \int_0^{+\infty} ds \  s N_L (s,H) 
=  L^a \int_{-\infty}^{+\infty} dH \int_0^{+\infty} ds \  s^{1-\tau} 
   \Phi \left( \frac{s}{L^{\rho}}, H s^{\psi}  \right) \nonumber \\
&& =  L^a \int_0^{+\infty} ds  \  s^{1-\tau-\psi} 
 \int_{-\infty}^{+\infty} dh   \phi \left( \frac{s}{L^{\rho}}, h  \right) 
\sim  L^a \left( L^{\rho} \right)^{2-\tau-\psi}
\label{sumrulefssaval}
\end{eqnarray}
i.e. one obtains the following relation between exponents
\begin{eqnarray}
d=a+\rho (2-\tau-\psi)
\label{sumruleaval}
\end{eqnarray}
The difference with the previous relation of Eq. \ref{relationchiaval} yields 
\begin{eqnarray}
\rho  = \frac{\theta}{\psi} = 2 \theta
\label{rhofss}
\end{eqnarray}
in agreement with the value measured for droplets (see Fig. \ref{figdropletnonMF} (b)),
 for zero-field avalanches (see Fig. \ref{figPzerosigma=0.4} (b)), and for integrated avalanches
(see Fig. \ref{figQintsigma=0.4} (b)).

The exponent $a$ governing the total number of avalanches (Eq. \ref{nltotla}) then reads
from Eq. \ref{relationchiaval}, Eq. \ref{rhofss} and $\tau=1+\frac{\theta}{d}$
 \begin{eqnarray}
a= \theta+d-\rho (2-\tau) = d -  \theta + 2 \frac{\theta^2}{d}
\label{aresult}
\end{eqnarray}

Besides our numerical checks concerning the Dyson hierarchical RFIM,
we may also compare with the numerical results of Ref. \cite{liu-dahmen}
concerning the statistics of equilibrium avalanches in the short-range RFIM in $d=3$ :
using the value of droplet exponent $\theta \simeq 1.49$ \cite{middleton3D},
we expect that zero-field avalanches correspond to the exponent $\tau = 1 + \frac{\theta}{d} \simeq 1.5$
(unfortunately very close to the mean-field value !), and that the integrated
 avalanches correspond to the exponent $\tau_{int}=3/2+ \frac{\theta}{d} \simeq 2. $,
in agreement with the value measured in Ref. \cite{liu-dahmen}.

In the non-mean-field region $d>d_c$ where usual finite-size scaling forms breaks down,
the analysis of finite-size properties requires a more subtle analysis, as discussed 
in the next section.

\section{ Finite-size properties in the 'Mean-field' region $d>d_c$  }

\label{sec_fsmf}

As is well known, the usual finite-size scaling properties are not valid in the mean-field region
$d>d_c$ (see for instance \cite{binder_nauenberg,aharony,luijten,jones,ahrens} and references therein).
In particular, the equalities of Eq. \ref{ratiosfss} concerning thermodynamic observables
are not valid anymore, because the correlation length $\xi \sim (W-W_c)^{-\nu}$
 is not the only important divergent length scale in the system.
These properties can be more clearly understood by the scaling theory developed by Coniglio
\cite{coniglio} on the example of the percolation transition for $d>d_c$. 
In section \ref{sec_mfdroplet} 
concerning the statistics of low-energy excitations in the mean-field region
$d>d_c$, we have already started to explain how Coniglio's approach could be adapted to the RFIM.
In the present section, we continue this analysis and derive the consequences for the finite-size
properties of various observables.

\subsection{ Finite-size properties at $H=0$ as a function of $(W-W_c)$  }

In section \ref{sec_mfdroplet} concerning the statistics of low-energy excitations in the mean-field region
$d>d_c$, we have recalled how the irrelevance of loops for $d>d_c$ leads to Eq. \ref{ns}
for the density of droplets, and to the fractal dimension $d_f=2/\nu_{MF}=2 \theta_{MF}$ (Eq. \ref{mfdf}).
 We have already described how
 the number of clusters $N_{\xi^d}$ in a correlated volume $\xi^d$
grows as $\xi^{d-d_c}$ (Eq. \ref{nxi}). This means that the correlation length $\xi$ is not the 
only important length scale (in contrast to the region $d<d_c$). 
In particular, it is clear that another important scale is the smaller
length $\xi_1$ defined by the requirement that
the number $N_{\xi_1^d}$ of clusters in a volume $\xi_1^d$ is one
\begin{eqnarray}
N_{\xi_1^d}  = \xi_1^d \sum_{s} n(s) = \xi_1^d \xi^{-d_c} =1 \ \ {\rm yields } \ \ 
\xi_1= \xi^{ \frac{d_c}{d}} = (W-W_c)^{- \nu_{MF} \frac{d_c}{d} }
\label{nxi1}
\end{eqnarray}
For instance in the short-range case where $\nu_{MF}=1/2 $ and $d_c=6$, one has
$\xi \sim (W-W_c)^{-1/2}$ and $\xi_1 \sim (W-W_c)^{-3/d}$.
In the one-dimensional ($d=1$) long-range case where $\nu_{MF}=1/\sigma $ and $d_c=3 \sigma$,
one has $\xi \sim (W-W_c)^{-1/\sigma}$ and $\xi_1 \sim (W-W_c)^{-3}$.

In a finite system of linear size $L$, one may thus expect three regimes :

(i) for $L<\xi_1$ : there exists $N_c(L)=1$ cluster of singular energy $E_c(L)=L^{\theta}$

(ii)  for $\xi_1<L<\xi$ : there exists $N_c(L)=\frac{L^d}{\xi_1^d}$ clusters
 of singular energy $E_c(L)=L^{\theta}$

(iii)  for $\xi<L$ : there exists $N_c(L)=\frac{L^d}{\xi_1^d}$ clusters
 of singular energy $E_c(L)=\xi^{\theta}$

The singular energy density $e_{sing}(L,W)$ will then present a complicated finite-size
form involving the two ratios $L/\xi_1$ and $\xi/L$
\begin{eqnarray}
e_{sing}(L,W,H=0) = \frac{N_c(L) E_c(L)}{L^d} = \frac{L^{\theta}}{\xi_1^d} 
\Phi \left( \frac{ \xi_1 }{ L } ;  \frac{ L }{\xi} \right)
\label{esingtworatios}
\end{eqnarray}
where the function $\Phi(x_1,x_2)$ should describe the crossover between (i) and (ii) 
when $x_2=0$
\begin{eqnarray}
\Phi (x_1=0,x_2=0) && =1 \nonumber \\
\Phi (x_1 \to + \infty ,x_2=0 ) && \simeq  x_1^d
\label{phi1}
\end{eqnarray}
and should describe the crossover between (ii) and (iii) when $x_1=0$
\begin{eqnarray}
\Phi (x_1=0, x_2 \to + \infty ) && \simeq  \frac{1}{x_2^{\theta} }
\label{phi2}
\end{eqnarray}

Note that in usual thermal transitions where there is no droplet exponent $\theta=0$, 
the crossover between (ii) and (iii) actually disappears, so that 
 the finite-size behaviors are only governed by the single ratio $L/\xi_1$ 
as proposed in  \cite{binder_nauenberg,aharony,luijten,jones,ahrens}.
However in the presence of the droplet exponent $\theta>0$ for the RFIM,
we expect that this simple recipe does not work anymore.
In particular, the finite-size behavior exactly at criticality $W_c$ (Regime (i) since
$\xi_1$ and $\xi$ are infinite)
\begin{eqnarray}
e_{sing}(L,W_c) = \frac{ L^{\theta}}{ L^{d}} = L^{\theta-d}
\label{esingmffinitesize}
\end{eqnarray}
is not connected to the result for $W \ne W_c$ in the thermodynamic limit $L \to +\infty$
(Regime (iii) since $\xi_1$ and $\xi$ are finite)
\begin{eqnarray}
e_{sing}(L=\infty,W \ne W_c) = \frac{\xi^{\theta}}{\xi_1^d} = \xi^{\theta-d_c}
 = \xi^{\frac{1}{\nu_{MF}}- \frac{3}{\nu_{MF}}} = \vert W-W_c \vert^2
\label{esingmf}
\end{eqnarray}
via a single crossover.

Let us now translate this scaling analysis for the finite-size properties exactly at $W_c$
 as a function of the external field $H$, to clarify the finite-size properties of the magnetization and 
of the susceptibility that can be obtained from the singularity of the energy density
by successive derivation with respect to $H$.

\subsection{ Finite-size properties at $W_c$ as a function of $H$  }

\begin{figure}[htbp]
  \includegraphics[height=6cm]{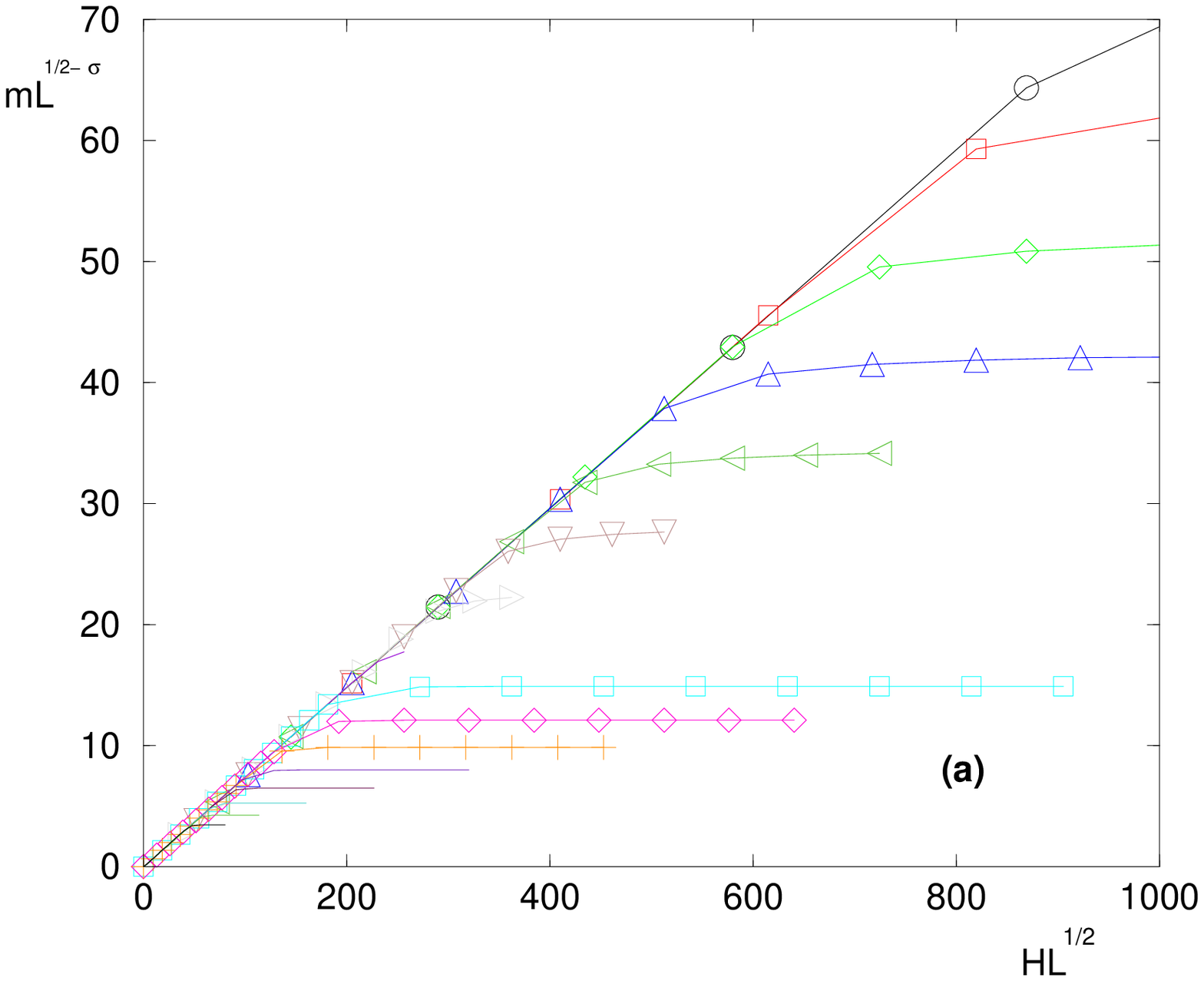}
\hspace{2cm}
 \includegraphics[height=6cm]{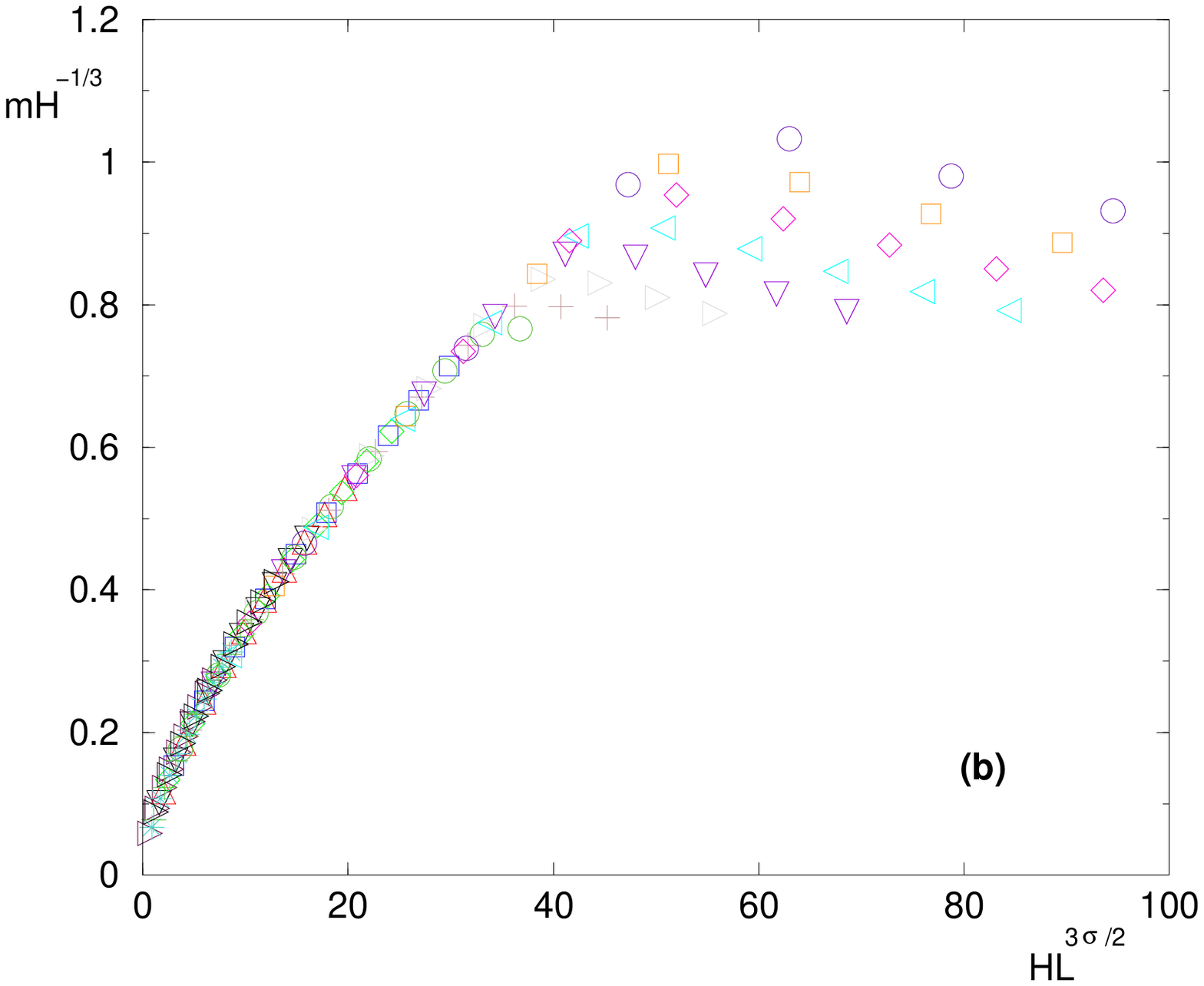}
\vspace{1cm}
\caption{ Finite-size scaling of the magnetization $m(H,L)$ at $W_c$
 as a function of the external magnetic field $H$ for $\sigma=0.2$ (mean-field region) : 
(a) $L^{1/2-\sigma} m(H,L)$ as a function of the scaling variable $H L^{1/2}$ (Eq. \ref{fss54xi1})
(b) $H^{-1/3} m(H,L)$ as a function of the scaling variable $H L^{3 \sigma/2}$ (Eq. \ref{fss54xi2}).  }
\label{figfss54sigma=0.2}
\end{figure}

In section \ref{sec_mfthermo}, we have described the mean-field critical exponents 
for thermodynamic observables at $H=0$. Let us now describe what happens as a function of
the external field $H$ exactly at $W_c$ in the thermodynamic limit $L \to +\infty$
\begin{eqnarray}
e_{sing}(W_c,H) && \sim  \vert H \vert^{\frac{1}{\delta_{MF}}+1} \ \nonumber \\
m(W_c,H) && \sim \vert H \vert^{\frac{1}{\delta_{MF}}}  \nonumber \\
\chi(W_c,H) && \sim  \vert H \vert^{\frac{1}{\delta_{MF}}-1} 
\label{thermoHmf}
\end{eqnarray}
in terms of the exponent 
\begin{eqnarray}
\delta_{MF}=  1+ \frac{\gamma_{MF}}{\beta_{MF}} =3 
\label{deltamf}
\end{eqnarray}
To reproduce the correct divergence of the susceptibility $\chi(W_c,H) \sim \vert H \vert^{-2/3}$ 
via the analog of Eq. \ref{ximf}, the analog of
Eq. \ref{correconnectedgaussian} for the Fourier transform of the connected correlation reads
\begin{eqnarray}
  {\hat C}_{connected}(q) \sim \frac{1}{\vert H \vert^{2/3} + \vert q \vert^{\sigma} }
\label{correconnectedgaussianH}
\end{eqnarray}
so that the exponential decay for $H \ne 0$ is governed by
\begin{eqnarray}
C_{connected}(r) \sim  \int dq e^{i q r} \frac{1}{\vert H \vert^{2/3}+ q^{\sigma} }
 \propto e^{- \frac{r}{\xi^{(H)}}}  \ \ {\rm with } \ \ \xi^{(H)} \sim \vert H \vert^{-\nu_{MF}^{(H)}} \ \ 
{\rm and } \ \ \nu_{MF}^{(H)}=\frac{2}{3\sigma}
\label{numfH}
\end{eqnarray}

The the density $n(s)$ of clusters of size $s$ per spin of Eq. \ref{ns} becomes
\begin{eqnarray}
n(s)  = \frac{1}{s^{\tau_{MF}+1}} e^{- s \vert H \vert^\frac{4}{3} } 
\ \ {\rm with } \ \ \tau_{MF}=\frac{3}{2}
\label{nsH}
\end{eqnarray}
In a volume $l^d$, the number $N_{l^d} $ of clusters has
 the following singularity in terms of the cut-off $s_{max} \sim  \vert H \vert^{-\frac{4}{3}}$ 
of Eq. \ref{nsH}
\begin{eqnarray}
N_{l^d}  = l^d \sum_{s} n(s) =l^d [ s_{max} ]^{- \tau_{MF}}
 = l^d \vert H \vert^{ \frac{4}{3}\tau_{MF}} = l^d \vert H \vert^{2}
\label{nlH}
\end{eqnarray}
So the length-scale $\xi_1$ introduced in Eq. \ref{nxi1} now diverges as
\begin{eqnarray}
\xi_1^{(H)}=  \vert H \vert^{ - \frac{2}{d} }
\label{nxi1H}
\end{eqnarray}

The analysis leading to Eq. \ref{esingtworatios} is now the same, provided
one use $ \xi_1^{(H)}$ and $ \xi^{(H)} \sim (\xi_1^{(H)} )^{d/d_c}= H^{-2/d_c}$,
 i.e. we may write after changes of variables
to make clearer the dependence upon $H$
\begin{eqnarray}
e_{sing}(L,W_c,H>0)  = L^{\theta} H^2  
{\cal E} \left( \frac{ 1 }{ H L^{d/2}   } ;  H  L^{d_c/2}  \right)
\label{esingtworatiosH}
\end{eqnarray}
where the function ${\cal E}(y_1,y_2)$ satisfies
\begin{eqnarray}
{\cal E} (y_1=0,y_2=0) && =1 \nonumber \\
{\cal E} (y_1 \to + \infty ,y_2=0 ) && \simeq  y_1^2  \nonumber \\
{\cal E} (y_1=0, y_2 \to + \infty ) && \simeq  y_2^{-\frac{2\theta}{d_c}} 
\label{psi2}
\end{eqnarray}
to reproduce the three regimes $L \ll \xi_1$, $\xi_1 \ll L \ll \xi$ and $ \xi \ll L$ 
\begin{eqnarray}
e_{sing}(L,W_c,H>0)  && \simeq   L^{\theta-d} \ \ \ {\rm for } \ \ L \ll H ^{ - 2/d }  \nonumber \\
e_{sing}(L,W_c,H>0)  && \simeq   L^{\theta} H^2 \ \ \ {\rm for } \ \  H ^{ - 2/d }\ll L \ll \xi  H^{-2/d_c} \nonumber \\
e_{sing}(L,W_c,H>0)  && \simeq   H^{2-2 \theta/d_c} = H^{4/3} \ \ \ {\rm for } \ \  H^{-2/d_c} \ll L 
\label{esingtworatiosH3regimes}
\end{eqnarray}

By differentiation with respect to $H$, we thus expect the following finite-size behavior for
the magnetization 
\begin{eqnarray}
m(L,W_c,H>0)  = L^{\theta} H {\cal M} \left( \frac{ 1 }{ H L^{d/2}   } ;  H  L^{d_c/2}  \right)
\label{mtworatiosH}
\end{eqnarray}
where the function ${\cal M} (y_1,y_2)$ satisfies
\begin{eqnarray}
{\cal M} (y_1=0,y_2=0) && =1 \nonumber \\
{\cal M} (y_1 \to + \infty ,y_2=0 ) && \simeq  y_1  \nonumber \\
 {\cal M} (y_1=0, y_2 \to + \infty ) && \simeq  y_2^{-\frac{2\theta}{d_c}} 
\label{Mscal2}
\end{eqnarray}
to reproduce the three regimes $L \ll \xi_1$, $\xi_1 \ll L \ll \xi$ and $ \xi \ll L$ 
\begin{eqnarray}
m (L,W_c,H>0)  && \simeq   L^{\theta-d/2} \ \ \ {\rm for } \ \ L \ll H ^{ - 2/d }  \nonumber \\
m (L,W_c,H>0)  && \simeq   L^{\theta} H \ \ \ {\rm for } \ \  H ^{ - 2/d }\ll L \ll \xi  H^{-2/d_c} \nonumber \\
m (L,W_c,H>0)  && \simeq    H^{1/3} \ \ \ {\rm for } \ \  H^{-2/d_c} \ll L 
\label{mtworatiosH3regimes}
\end{eqnarray}

To test the presence of the two crossovers of Eq. \ref{mtworatiosH3regimes} for the magnetization,
we have plotted our data for the Dyson hierarchical model for $\sigma=0.2$
in terms of the ratios $L/\xi_1$ and $L/\xi$ respectively.
On  Fig. \ref{figfss54sigma=0.2} (a), we show the test of the scaling form involving $L/\xi_1 $
\begin{eqnarray}
 L^{1/2-\sigma} m (L,W_c,H>0)  = {\cal M}_1 \left( H L^{\frac{1}{2}} \right) 
\label{fss54xi1}
\end{eqnarray}
whereas on Fig. \ref{figfss54sigma=0.2} (b), we show  the test of the other scaling form
involving $L/\xi $
\begin{eqnarray}
H^{-1/3} m (L,W_c,H>0)  = {\cal M}_2 \left( H L^{\frac{3 \sigma}{2}} \right) 
\label{fss54xi2}
\end{eqnarray}

By another differentiation of the magnetization of Eq. \ref{mtworatiosH3regimes}
with respect to $H$, one obtains that the crossover at $\xi_1$
actually disappears in the finite-size behavior of the susceptibility
(i.e. the regimes (i) and (ii) give the same contribution)
\begin{eqnarray}
\chi(L,W_c,H>0)  = L^{\theta}   G \left(   H  L^{d_c/2}  \right)
\label{chitworatiosH}
\end{eqnarray}
where the function $G(y)$ satisfies
\begin{eqnarray}
G(y=0) && =1 \nonumber \\
G ( y_2 \to + \infty ) && \simeq  y^{-\frac{2\theta}{d_c}} 
\label{chiscal2}
\end{eqnarray}
corresponding to the two regimes $ L \ll \xi$ and $ \xi \ll L$ 
\begin{eqnarray}
\chi (L,W_c,H>0)  && \simeq   L^{\theta} \ \ \ {\rm for } \ \ L \ll \xi  H^{-2/d_c} \nonumber \\
\chi (L,W_c,H>0)  && \simeq    H^{-2/3} \ \ \ {\rm for } \ \  H^{-2/d_c} \ll L 
\label{chitworatiosH3regimes}
\end{eqnarray}

\section{ Conclusion }

\label{sec_conclusion}

In this paper, we have discussed the statistics of 
low-energy excitations and of equilibrium avalanches for  
the zero-temperature finite-disorder critical point of the Random-field Ising model (RFIM).
Besides its role is the thermodynamics, the droplet exponent $\theta$ is expected to govern
the distribution $dl/l^{1+\theta}$ of the linear-size $l$
 of low-energy excitations or zero-field avalanches.
In terms of the number $s \sim l^{d_f}$ of spins of these excitations of fractal dimension $d_f$,
the power-law distribution thus reads $ds/s^{1+\theta/d_f}$.
 In the non-mean-field region $d<d_c$, droplets are compact $d_f=d$, 
whereas in the mean-field region $d>d_c$, droplets have a fractal dimension $d_f=2 \theta$
leading to the well-known mean-field result $ds/s^{3/2}$. We have also discussed 
in details finite-size effects, both in the non-mean-field region $d<d_c$ 
where standard finite-size scaling is valid, and in the mean-field region $d>d_c$,
where standard finite-size scaling breaks down, because the correlation length
is not the only relevant length scale. We have explained how to adapt the Coniglio's scaling
approach to understand the finite size properties of the RFIM 
for $d>d_c$ in terms of the two scales $\xi_1$ and $\xi$.  
All expectations have been checked numerically for the Dyson hierarchical version of the RFIM,
where large systems can be studied with a good statistics via exact recursion,
and where the droplet exponent $\theta$ can be varied as a function of the parameter $\sigma$ of the effective power-law ferromagnetic coupling.

 \appendix

\section{ Exact recursion for the Dyson hierarchical RFIM }

\label{app_recursion}

In this Appendix, we explain how the sequence of ground states 
that occur as a function of the external field $H$ in any given disordered sample
can be easily computed by recursion for the Dyson model introduced in Eq. \ref{hamilton}.
 All numerical results presented in the text have been obtained by this method.

\subsection{ Exact recursion for the partition function }

Following the method of \cite{Kim77} for the ferromagnetic model,
we use the Gaussian identity
\begin{eqnarray}
e^{ \left[\beta J_N \left(\sum_{i=1}^{2^N} S_i \right)^2 \right]}
= \sqrt{ \frac{\beta}{\pi} } \int_{-\infty}^{+\infty} dx
e^{- \beta x^2 + 2 \beta  x \sqrt{ J_N} \left(\sum_{i=1}^{2^N} S_i \right) }  
\label{gaussian_identity}
\end{eqnarray}
to rewrite
the partition function for a system containing $2^N$ spins
of Hamiltonian given by Eq \ref{hamilton}
\begin{eqnarray}
Z_{2^N}(\beta;H;\{h_1,...,h_{2^N}\}) \equiv \sum_{S_1 = \pm 1 ; ..., S_{2^N}=\pm 1}
e^{- \beta {\cal H}_{2^N}(H;\{h_1,...,h_{2^N}\}) } 
\label{zdef}
\end{eqnarray}
 as
\begin{eqnarray}
&& Z_{2^N}(\beta;H;\{h_1,...,h_{2^N}\}) 
 =  \sqrt{ \frac{\beta}{\pi} } \int_{-\infty}^{+\infty} dx  
e^{- \beta x^2} \nonumber \\
&& \times Z_{2^{N-1}}(\beta;H+ 2 x \sqrt{ J_N} ;\{h_1,...,h_{2^{N-1}}\}) 
\times  Z_{2^{N-1}}(\beta;H+ 2 x \sqrt{ J_N} ;\{h_{2^{N-1}+1},...,h_{2^{N}}\}) 
\label{zrec}
\end{eqnarray}
On the right hand-side appear the partition functions of the two half-systems
in the modified exterior magnetic field $H'=H+ 2 x \sqrt{ J_N} $.

The initial condition corresponding to $N=0$ and a single spin $2^0=1$ reads
\begin{eqnarray}
Z_{1}(\beta;H;\{h_1\}) 
&& = \sum_{S_1 = \pm 1} e^{ \beta(H+h_1) S_1} = 2 \cosh \left( \beta(H+h_1) \right)
\label{zini}
\end{eqnarray}

\subsection{ Limit of zero-temperature }

In the limit of zero-temperature where $\beta=1/T \to +\infty$, 
the partition function of Eq. \ref{zdef} becomes dominated
 by the ground-state energy $E^{GS}_{2^N}(H;\{h_1,...,h_{2^N}\})$
\begin{eqnarray}
Z_{2^N}(\beta;H;\{h_1,...,h_{2^N}\}) \opsimeq_{\beta \to +\infty}
e^{- \beta E^{GS}_{2^N}(H;\{h_1,...,h_{2^N}\}) } 
\label{egs}
\end{eqnarray}

In the limit $\beta \to +\infty$, the recursion of Eq. \ref{zrec} 
can be thus evaluated via the saddle-point approximation.
This yields the following recursion for the ground state energy
\begin{eqnarray}
&&   E^{GS}_{2^N}(H;\{h_1,...,h_{2^N}\}) \nonumber \\
&& = {\rm min}_{x} 
 \left[    x^2  
+ E^{GS}_{2^{N-1}}(H+ 2 x \sqrt{ J_N} ;\{h_1,...,h_{2^{N-1}}\}) 
+  E^{GS}_{2^{N-1}}(H+ 2 x \sqrt{ J_N} ;\{h_{2^{N-1}+1},...,h_{2^{N}}\}) \right] 
 \nonumber \\
&& = {\rm min}_{H'} \left[     \frac{(H'-H)^2}{4 J_N}   
+ E^{GS}_{2^{N-1}}(H';\{h_1,...,h_{2^{N-1}}\}) 
+  E^{GS}_{2^{N-1}}(H' ;\{h_{2^{N-1}+1},...,h_{2^{N}}\}) \right] 
\label{erec}
\end{eqnarray}

The initial condition (Eq. \ref{zini}) reads
\begin{eqnarray}
E_{1}(H;\{h_1\})  = - \vert H+h_1 \vert
\label{eini}
\end{eqnarray}

\subsection{ Sequence of ground states  as a function of the exterior field $H$ }

It is convenient to characterize each disordered sample of $2^n$ spins
by its 'sequence' of ground states as the exterior field $H$ is swept from $(-\infty)$
to $(+\infty)$ : this sequence contains a certain number $(1+p_{max})$ of configurations
$\{C_0, C_1, ...,  C_{p_{max}}\}$, where 
$C_0$ is the configuration where all spins are negative $S_i=-1$ (ground state when $H \to -\infty$), 
and where $C_{p_{max}}$ is the configuration where all spins are 
positive $S_i=1$ (ground state when $H \to +\infty$).

The energy of each configuration $C_{p}$ depends linearly on the exterior field $H$,
with a slope determined
 by the magnetization $M_{C_{p}}=\sum_i S_i$ of the configuration $C_{p}$
\begin{eqnarray}
E(C_{p},H) = - M_{C_{p}} H + a_{C_p}
\label{ecih}
\end{eqnarray}
The value of the field  $H=H_{C_{p},C_{p+1}}$
where the ground state changes from $C_{p}$ to $C_{p+1}$
is simply given by the intersection of the corresponding two lines (Eq. \ref{ecih})
\begin{eqnarray}
 H_{C_{p},C_{p+1}}  = \frac{a_{C_{p+1}}-a_{C_p}}{(m_{C_{p+1}}- m_{C_{p}})}
\label{seuilhcp}
\end{eqnarray}

\subsection{ Notion of 'no-passing rule'  }

The notion of 'no-passing rule' has been first developed for non-equilibrium
dynamics concerning charge-density waves \cite{middletonCDW}, driven elastic manifolds
\cite{rosso_nop} and the non-equilibrium dynamics of the RFIM \cite{noneqrfim_nop}. 
This notion has been then extended to the equilibrium of the RFIM \cite{frontera,liu_nopassing},
where it means that the sequence of ground states that appear as a function of 
the external field $H$ are 'ordered' in magnetization
\begin{eqnarray}
M_{C_{0}}=-2^n < M_{C_{1}} < M_{C_{2}} < .. < M_{C_{p_{max}}} =2^n
\label{ordermi}
\end{eqnarray}
Since the difference between the magnetizations of two consecutive configurations
is bounded from below by the value $2$ that corresponds to a single spin-flip
\begin{eqnarray}
m_{C_{p+1}} - m_{C_{p}} \geq 2
\label{ordermi2}
\end{eqnarray}
so that the number $p_{max}$ is bounded by the number of spins
\begin{eqnarray}
1 \leq p_{max} \leq 2^n
\label{ncmax}
\end{eqnarray}
i.e. it always remain very small with respect to the total number $2^{2^n}$
of possible configurations.

\subsection{ Recursion on the sequence of  ground states  as a function of the exterior field $H$  }

Let us now assume that we know the sequences of ground states in $H$
 of two independent half-systems of size $2^{N-1}$, and we wish 
to construct the sequence for the whole system when these two half-systems are coupled.

We consider the set of pairs of ground states of the two subsystems that exist
at a given same exterior field $H'$ :
let us call $I_{p_1,p_2}$ the interval of the exterior field $H'$, 
where the first subsystem has for ground state
the configuration $C_{p_1}^{(1)}$ of energy (Eq. \ref{ecih})
\begin{eqnarray}
E_{2^{N-1}}^{(1)}(H') =  - m_{C^{(1)}_{p_1}} H' + a^{(1)}_{p_1}
\label{esub1}
\end{eqnarray}
and where the second subsystem has for ground state
the configuration $C_{p_2}^{(2)}$ of energy 
\begin{eqnarray}
E_{2^{N-1}}^{(2)}(H') =  - m_{C^{(2)}_{p_2}} H' + a^{(2)}_{p_2}
\label{esub2}
\end{eqnarray}
The recursion of Eq. \ref{erec} means that in this interval $I_{p_1,p_2}$ of $H'$,
the function that has to be minimized reads
\begin{eqnarray}
\phi_{I_{p_1,p_2}}(H') && =     \frac{(H'-H)^2}{4 J_N} + E_{2^{N-1}}^{(1)}(H')
+  E_{2^{N-1}}^{(2)}(H') 
\nonumber \\
&& =     \frac{(H'-H)^2}{4 J_N}   
 - m_{C^{(1)}_{p_1}} H' + a^{(1)}_{p_1}  - m_{C^{(2)}_{p_2}} H' + a^{(2)}_{p_2}
\label{phihprime}
\end{eqnarray}
The minimization over $H'$ 
\begin{eqnarray}
0= \partial_{H'} \phi_{I_{p_1,p_2}}(H') && =     \frac{ (H'-H)}{2 J_N}
 - m_{C^{(1)}_{p_1}}   - m_{C^{(2)}_{p_2}} 
\label{phihprimederi}
\end{eqnarray}
yields the solution
\begin{eqnarray}
H'_*(H) = H+2 J_N ( m_{C^{(1)}_{p_1}}  + m_{C^{(2)}_{p_2}})
\label{soluhprime}
\end{eqnarray}
that corresponds to the value (Eq \ref{phihprime})
\begin{eqnarray}
\phi_{I_{p_1,p_2}}(H'_*(H)) &&  =  
  J_N ( m_{C^{(1)}_{p_1}}  + m_{C^{(2)}_{p_2}})^2  
 - (m_{C^{(1)}_{p_1}}+  m_{C^{(2)}_{p_2}})( H+2 J_N ( m_{C^{(1)}_{p_1}}  + m_{C^{(2)}_{p_2}}) )+ a^{(1)}_{C_{p_1}} + a^{(2)}_{C_{p_2}}\nonumber \\
&&  =  
     - J_N ( m_{C^{(1)}_{p_1}}  + m_{C^{(2)}_{p_2}})^2 
 - (m_{C^{(1)}_{p_1}}+  m_{C^{(2)}_{p_2}}) H+ a^{(1)}_{p_1} + a^{(2)}_{p_2}
\nonumber \\
&& \equiv E_{2^N} (C=(C^{(1)}_{p_1},C^{(2)}_{p_2},H)
\label{phihprimep1p2}
\end{eqnarray}
representing the energy of the global configuration $C=(C^{(1)}_{p_1},C^{(2)}_{p_2})$
made of $C^{(1)}_{p_1} $ and $C^{(2)}_{p_2} $ for the two sub-systems,
when the exterior field is $H$.

We now need to minimize over all possible intervals $I_{p_1,p_2}$ for $H'$ (Eq. \ref{erec})
\begin{eqnarray}
   E^{GS}_{2^N}(H) 
  = {\rm min}_{(C^{(1)}_{p_1},C^{(2)}_{p_2})} \left[ E_{2^N} (C=(C^{(1)}_{p_1},C^{(2)}_{p_2},H) \right]
\label{eminp1p2}
\end{eqnarray}
where the minimization is over all the pairs $(C^{(1)}_{p_1},C^{(2)}_{p_2})$ 
of configurations that are ground-states of the isolated 
 sub-systems for some same value $H'$ of the exterior field.

In practice, we have thus used the following procedure :

(i)  We make the ordered list in the field $H'$ of the pairs
  $(C^{(1)}_{p_1},C^{(2)}_{p_2})$
of configurations that are ground-states of the isolated two sub-systems.
Let us we call $\{C_0,C_1,...,C_{q_{max}}\}$ these 'candidate'
configurations of the whole system, and compute their energies.
For instance if $C_q=(C^{(1)}_{p_1},C^{(2)}_{p_2})$, its energy is simply
\begin{eqnarray}
 E_{2^N} (C_q=(C^{(1)}_{p_1},C^{(2)}_{p_2},H) && = - M_{C_q} H +a_{C_q}
\nonumber \\
 M_{C_q} && \equiv   M_{C^{(1)}_{p_1}}  + M_{C^{(2)}_{p_2}}
+ a^{(1)}_{C_{p_1}} + a^{(2)}_{C_{p_2}}  \nonumber \\
 a_{C_q} && \equiv  - J_N ( M_{C^{(1)}_{p_1}}  + M_{C^{(2)}_{p_2}})^2 
+ a^{(1)}_{C_{p_1}} + a^{(2)}_{C_{p_2}}
\label{ep1p2new}
\end{eqnarray}

(ii) Now these energies are in competition to be the ground state of the whole system
at some given exterior field $H$. Since they are ordered in magnetization,
 one may proceed as follows \cite{frontera,liu_nopassing}.
We start from the two known extremal configurations :
the first configuration is $C_0=(C^{(1)}_{0},C^{(2)}_{0})$ where all spins are negative
(ground-state for $H \to -\infty$) and the last configuration is 
$C_{last}=(C^{(1)}_{p_1^{max}},C^{(2)}_{p_2^{max}})$ where all spins are positive
(ground-state for $H \to +\infty$). We compute the crossing field $H_{0,last}$
where the two corresponding energies cross. We now compute the energies of all intermediate 
candidates at this crossing field $H_{0,last}$ and select the minimal value :
the corresponding configuration $C_{q_a}$ is then the ground state at $H_{0,last}$.
We may now iterate this procedure : we compute the crossing field $H_{0,q_a}$
and find the minimal energy at this field among the candidates $0 \leq q \leq q_a$;
similarly we compute the crossing field $H_{q_a,last}$ etc...
This method allows to compute the sequence of ground states that really appear
as a function of $H$ for the whole interval.
To make even clearer this procedure, we now describe as an example
the first step where two systems of one spin
are coupled to form a system of 2 spins.

\subsection{ Example with $N=1$ corresponding to $2^1=2$ spins }

Each subsystem contains only one spin. 
So the sequence of the ground state of the first subsystem
as function of the exterior field 
contains only the two configurations $(C_0^{(1)},C_1^{(1)})$ 
corresponding to $(S_1=-1,S_1=+1)$ and the energies read (Eq.  \ref{eini} and \ref{ecih}  )
\begin{eqnarray}
E(C_{0}^{(1)},H') && = - m_{C_{0}^{(1)}} H' + a_{C_0^{(1)}} = H' +h_1 \nonumber \\
E(C_{1}^{(1)},H') && = - m_{C_{1}^{(1)}} H' + a_{C_1^{(1)}} = - H' - h_1
\label{ecih1}
\end{eqnarray}
so that the field $H'= H_{C_{0}^{(1)},C_{1}^{(1)}}$ where the ground state changes from
$C_{0}^{(1)} $ to $C_{1}^{(1)} $ is
\begin{eqnarray}
H_{C_{0}^{(1)},C_{1}^{(1)}}=-h_1
\label{ecih1flip}
\end{eqnarray}

Similarly, the sequence of the second sub-system is described by the parameters
\begin{eqnarray}
E(C_{0}^{(2)},H') && = - m_{C_{0}^{(2)}} H' + a_{C_0^{(2)}} = H' +h_2 \nonumber \\
E(C_{1}^{(2)},H') && = - m_{C_{1}^{(2)}} H' + a_{C_1^{(2)}} = - H' - h_2 \nonumber \\
H_{C_{0}^{(2)},C_{1}^{(2)}} && =-h_2
\label{ecih2}
\end{eqnarray}

Let us assume that $h_1 > h_2$ (otherwise exchange the labels 1 and 2).
Then $H_{C_{0}^{(1)},C_{1}^{(1)}}=-h_1 < -h_2= H_{C_{0}^{(2)},C_{1}^{(2)}}$
so that the ordered list of candidates for the ground states of the whole system
 of two spins is $(C_0=(--);C_1=(+-);C_2=(++)$.
The energies of these three candidates are
\begin{eqnarray}
E(C_0,H) && =  2H + h_1+h_2 - 4 J_1    \nonumber \\
E(C_1,H) && =  - h_1+h_2   \nonumber \\
E(C_2,H) && =  -2H - h_1-h_2 - 4 J_1 
\label{ecandidates}
\end{eqnarray}
The crossing field  $H_{C_0,C_2}$ where the two energies of the extremal ground states cross reads
\begin{eqnarray}
H_{C_{0},C_{2}}   =   -  \frac{h_1+h_2}{2} 
\label{hcrossing}
\end{eqnarray}
and the corresponding energy reads
\begin{eqnarray}
E(C_0,H_{C_{0},C_{2}}) = E(C_2,H_{C_{0},C_{2}}) =  - 4 J_1
\label{ecrossing}
\end{eqnarray}
We now have to compute the energy of the intermediate candidate $C_1$ at this crossing field
\begin{eqnarray}
E(C_1,H_{C_{0},C_{2}}) =  - h_1+h_2  
\label{ec1crossing}
\end{eqnarray}
to see if it is lower or bigger than the crossing energy of Eq. \ref{ec1crossing} :

(i) if $E(C_1,H_{C_{0},C_{2}}) <  E(C_0,H_{C_{0},C_{2}}) = E(C_2,H_{C_{0},C_{2}})$,
 then $C_1$ is indeed the true ground state at the crossing field $ H_{C_{0},C_{2}}$.
So the sequence of ground states
contains indeed the three configurations
$(C_0,C_1,C_2)$ with the frontiers given by the crossing fields $  H_{C_{0},C_{1}}$
and $H_{C_{1},C_{2}}$.

(ii) if $E(C_1,H_{C_{0},C_{2}}) >  E(C_0,H_{C_{0},C_{2}}) = E(C_2,H_{C_{0},C_{2}}) $, 
then the candidate $C_1$ has to be eliminated.
 The sequence of ground states only
contains the two configurations $(C_0=(--),C_2={++})$ with a frontier given by $H_{C_{0},C_{2}} $.
Note that the condition for this direct jump between the two ferromagnetic configurations reads
\begin{eqnarray}
 4 J_1 >    h_1 -h_2 (>0)
\label{directjump}
\end{eqnarray}
i.e. it is more probable at low disorder as it should.

\end{document}